\documentclass{JHEP3}
\usepackage{epsfig}

\newcommand{\be}{\begin{equation}}
\newcommand{\ee}{\end{equation}}
\newcommand{\bea}{\begin{eqnarray}}
\newcommand{\eea}{\end{eqnarray}}
\newcommand{\IR}{{\bf R}} \newcommand{\IC}{{\bf C}}

\newcommand{\SLt}[1]{SL(2,#1)}

\newcommand{\Slash}{\hspace{-2.2mm} /}
\newcommand{\Gsla}{G \hspace{-2.2mm} /^{(1)}}
\newcommand{\bGsla}{G \hspace{-2.2mm} /^{*(1)}}
\newcommand{\non}{\nonumber \\}
\newcommand{\sfv}{${\bf S}^5~$}
\def\obst{\delta {\tilde \psi}}

\newcommand{\h}[1]{{\hat #1}} \newcommand{\tl}[1]{{\tilde #1}}
\def\room{~\rule[-2mm]{0mm}{8mm}}
\def\ten{{\bf 10}} \def\eight{{\bf 8}}

\def\IZ{\relax\ifmmode\hbox{Z\kern-.4em Z}\else{Z\kern-.4em Z}\fi}

\def\bz{{\bar z}} \def\bh{{\bar h}}
\def\bi{{\bar i}}
\def\bj{{\bar j}} 

\def\gammad{{\gamma^\dag}} \def\gammar{{\gamma_r}} \def\Gammar{{\Gamma_r}}
\def\hM{{\hat M}} \def\hN{{\hat N}}
 
\def\hmu{{\hat \mu}} \def\hnu{{\hat \nu}}
\def\tOmega{\widetilde{\Omega}} \def\tomega{\widetilde{\omega}}
\def\te{{\widetilde e}} \def\tGamma{{\widetilde \Gamma}}
\def\cT{{\cal T}}
\def\del{\partial}
\def\hphi{K} 
\def\fpar{f_\parallel} \def\fperp{f_\perp}

\def\half{{1 \over 2}}

\def\tr{{\rm tr}}
\def\del{{\partial}}
\def\fsl{{F\Slash}}
\def\gsl{{G\Slash}}

\def\al{\alpha} \def\bt{\beta}
\def\gm{\gamma}  \def\eps{\epsilon}
\def\hal{{\hat \alpha}} \def\hbt{{\hat \beta}}
\def\hgm{{\hat \gamma}} 
\def\Llambda{\mbox{{\Large $\lambda$}}}

\def\cn{{\cal N}} \def\cM{{\cal M}}
\newcommand{\sbsection}[1]{\vspace{.5cm} \noindent {\it #1}}
\def\ads{AdS_5 \times S^5}
\def\diff{{(\del \Slash + {2\over r} \gamma_r)}}

\preprint{WIS/17/02-MAY-DPP \\ TAUP-2703-02 \\ {\tt
hep-th/0205090}}

\title{On Exactly
Marginal Deformations of $\cn=4$ SYM and Type IIB Supergravity on
$AdS_5\times S^5$}

\author{Ofer Aharony\footnote{Incumbent of the Joseph and Celia Reskin
career development chair.}\\
Department of Particle Physics, The Weizmann Institute of Science \\
Rehovot 76100, Israel \\
\email{Ofer.Aharony@weizmann.ac.il} }
\author{Barak Kol
\\
School of Natural Sciences, Institute for Advanced Study \\
Einstein Drive, Princeton NJ 08540,
USA\\
\email{barak@sns.ias.edu} }
\author{Shimon Yankielowicz
\\
School of Physics and Astronomy \\
Beverly and Raymond Sackler Faculty of Exact Sciences \\
Tel Aviv University, Tel Aviv 69978,
Israel\\
\email{shimonya@post.tau.ac.il} }

\abstract{$\cn=4$ supersymmetric Yang-Mills theory with gauge
group $SU(N)$ ($N \geq 3$) is believed to have two exactly
marginal deformations which break the supersymmetry to $\cn=1$.
We discuss the construction of the string theory dual to these
deformations, in the supergravity approximation, in a perturbation series
around the $AdS_5\times S^5$ solution. We construct explicitly the deformed
solution at second order in the deformation. We show that
deformations which are marginal but not exactly marginal lead to
a non-conformal solution with a logarithmically running coupling
constant. Surprisingly, at third order in the deformation we find
the same beta functions for the couplings in field theory and in
supergravity, suggesting that the leading order beta functions
(or anomalous dimensions) do not depend on the gauge coupling
(the coefficient is not renormalized).}


\begin{document}

\section{Introduction}

A conformal field theory may have exactly marginal deformations
which preserve the conformal symmetry. The AdS/CFT correspondence
\cite{Maldacena:1998re,Gubser:1998bc,Witten:1998qj,Aharony:1999ti}
maps such deformations to continuous deformations of AdS
solutions of gravitational theories which preserve the conformal
isometries.

The AdS/CFT correspondence and its generalizations to
non-conformal theories provide an equivalence between
gravitational theories and non-gravitational theories.  In the
AdS case the non-gravitational theories are simply local
conformal field theories. Computations which can be performed on
both sides of the correspondence have provided many tests of the
equivalence between the two sides. The main utility of the
correspondence so far has been to learn about strongly coupled
non-gravitational theories from gravitational solutions (in
particular, from solutions of classical (super)gravity). In this
paper we will go in the other direction. We will analyze an issue
that is well-understood on the field theory side, and attempt to
use it to learn about gravitational solutions and to test the
AdS/CFT correspondence (in case there are still unconvinced
skeptics).

One general property of supersymmetric conformal field theories is
that many of them have exactly marginal deformations, preserving
superconformal invariance (see the paper by Leigh and Strassler
 \cite{Leigh:1995ep} and references therein). In
order to find a conformal theory one has to solve the constraints
for the vanishing of all the beta functions. Since there are as
many beta functions as there are couplings, these equations
generically will not have a non-zero solution (and if they do,
any additional solutions are expected to be isolated). However,
in $\cn =1$ supersymmetric field theories in four dimensions (or,
more generally, in theories with at least four supercharges) the
beta functions are linearly dependent on the gamma functions (the
anomalous dimensions). Thus, if for some reason there are fewer
gamma functions appearing in the beta function equations than
there are couplings, then there will generically exist a manifold
of solutions, with a (complex) dimension given by the difference
between the number of couplings and the number of equations.

We denote the space of conformal theories on which a specific
theory $X$ lies by $\cM_c(X)$, to be distinguished from the
moduli space of vacua $\cM$. Given the central role of $\cn=4$
supersymmetric Yang-Mills (SYM) theory among gauge theories it is
especially interesting to study $\cM_c(\cn=4)$.  Using the
methods described above, it was found \cite{Leigh:1995ep} that
$\cn =4$ SYM with gauge group $SU(N)$, $N \ge 3$ has, in addition
to the complex gauge coupling (which is exactly marginal and
preserves the full $\cn =4$ supersymmetry), two extra complex
exactly marginal deformations preserving $\cn=1$ supersymmetry.
Thus, for $SU(N)$ gauge groups with $N \geq 3$,
$\mbox{dim}(\cM_c(\cn =4\ SYM))=2_\IC+1_\IC$. This analysis was
carried out at weak coupling (using concepts like the anomalous
dimension of the elementary fields), but as the gauge coupling is
exactly marginal we do not expect the dimension of $\cM_c$ to
change as a function of the coupling, so we expect to find the
same dimension also at strong coupling.

The AdS/CFT correspondence maps conformal invariance in field
theory to $SO(4,2)$ isometries in the solution to some
gravitational theory, and exactly marginal deformations are
mapped to continuous families of solutions with $SO(4,2)$
isometries. Our objective is to understand the existence of such
families of solutions from the gravity side. In particular, since
$\cn=4$ SYM with $SU(N)$ gauge groups is believed to be
equivalent to type IIB string theory on $AdS_5\times S^5$, we
should have $\cM_c(\cn=4\ SYM)=\cM(IIB\ \mbox{ on } AdS^5 \times
S^5)$, and the $AdS_5\times S^5$ solution should belong to a
3-complex-dimensional family of continuous solutions (one of the
complex dimensions corresponds to changing the dilaton and axion,
but the other two are not known). In supersymmetric flat
compactifications of string theory one often finds such large
``moduli spaces'' of solutions, parameterized for instance by
geometric properties of Calabi-Yau manifolds. However, for AdS
compactifications the appearance of such ``moduli spaces'' is
less understood, and it seems that the only continuous
deformations which appear in the literature involve either the
string coupling constant or the integral of $p$-form potentials on
compact $p$-cycles.

For large $N$, weak coupling and strong 't Hooft coupling (large
$g_{YM}^2 N$), supergravity is expected to be a good
approximation, and we can study this problem in the supergravity
approximation, and look for a continuous family of type IIB
supergravity solutions\footnote{Not all deformations of conformal
field theories map to supergravity modes \cite{Aharony:2001pa},
but the deformations which we will discuss here are mapped to
supergravity modes so we can use supergravity to analyze them.}.
At least for small deformations away from the $AdS_5\times S^5$
solution we expect that supergravity will still be a good
approximation, so that close to the $\cn=4$ fixed line we should
have $\cM_c(\cn=4,\,N \to \infty, \lambda_{YM} \gg
1)=\cM^{SUGRA}(IIB\ \mbox{ on }\ AdS^5 \times S^5)$, where
$\lambda_{YM}\equiv g_{YM}^2 N$. We will thus construct a family
of type IIB supergravity backgrounds, which are expected to also
be good type IIB string theory backgrounds; it would be
interesting to construct the relevant string theories directly,
and perhaps to understand our results from the worldsheet point
of view.

Ideally, we would like to directly translate the field theory
arguments for the existence of exactly marginal deformations to
gravity, and to be able to prove their existence also on the
gravity side. However, it is not clear how to perform this
translation, for instance because we do not know how to identify
the anomalous dimensions of the elementary fields on the gravity
side (this is problematic because these are not gauge-invariant
objects, and they depend on how the fields are normalized). Thus,
it is interesting to understand the mechanism that guarantees
conformal deformations in gravity, and its relation to the field
theory mechanism. We were not able to do this, but we construct
an explicit solution for the deformation up to second order in the
perturbation, and we discuss the obstruction to conformal
invariance on the gravity side. We show that this obstruction has
the same form as the obstruction in field theory, coming from the
anomalous dimensions,
 but we have not yet been able to
make this relation precise.  We can also compute the deviation
from conformal invariance when we deform by couplings that are
not exactly marginal, namely the beta function. We find that at
leading order in the superpotential couplings, this deviation is
the same at strong coupling (using supergravity (SUGRA)) and at
weak coupling (using perturbation theory). We hope that our
results will be useful for a general understanding of exactly
marginal deformations in gravitational backgrounds. We are
confident that the methods we describe here will allow the
analysis of geometries other than $AdS_5 \times S^5$, in
particular $AdS_4 \times S^7$ where the number of the
deformations is not known and moreover the field theory side is
poorly understood.

Seen in a wider context this is a study of a certain space of
conformal theories. Exact descriptions of moduli spaces of {\it
vacua} $\cM(X)$, have become numerous in recent years with the
improved non-perturbative understanding of supersymmetric
theories. Given a conformal theory $X$, determining $\cM_c(X)$ is
similar to a ``moduli'' problem, and it is a natural and
fundamental question. Actually this is more than an analogy -- the
AdS/CFT correspondence maps the space of conformal deformations
to the moduli space of gravitational solutions on $AdS$, and also
in the case of perturbative string theory, the space of conformal
worldsheet theories maps to the moduli space of the space-time
theory. Here we make a first step in studying $\cM_c$, and we
believe that with our current understanding of exact moduli
spaces the tools are available to determine at least some
$\cM_c$'s completely.

Our problem could have been studied now for some years since the
discovery of the AdS-CFT correspondence\footnote{In fact, except
for the comparison with field theory, all our computations could
have been done already in the 1980's, though it would have been
hard to motivate them.}, and indeed it was studied to low orders
in the deformation in \cite{unpublished,Girardello:1998pd}.
However, it has resisted solution so far\footnote{A solution to
this problem was presented in \cite{Fayyazuddin:2002vh}, but it
seems to be singular and is not related to the solutions we
construct.}.  It has turned out to be easier to study
supersymmetric mass deformations of $\cn=4$ SYM (starting from
\cite{Girardello:1999bd,PolchinskiStrassler}), partly because
these can be analyzed using the truncation to 5d supergravity
(which does not include the deformations we are interested in
here). Gra\~ na and Polchinski \cite{GranaPolchinski} studied the
first order deformation in the $\ads$ solution
 for an arbitrary
superpotential, a result which we use and extend to higher orders.

\sbsection{Method and summary of results}

We start by analyzing the field theory in section \ref{FT}. Using
the form of the 1-loop gamma function at weak coupling and the
analysis of \cite{Leigh:1995ep} we write down an equation for the
exactly marginal deformations which is invariant under the
$SU(3)$ global symmetry group. This equation is quadratic in the
couplings, and we expect to find superconformal solutions if and
only if it is satisfied.

We turn to supergravity in section \ref{methods}. We take our
ansatz to be the most general ansatz with $SO(4,2)$ isometry, and
our equations are the supersymmetry (SUSY) variation equations
rather than the equations of motion of type IIB since we insist
on a supersymmetric background. Actually we require only
invariance under four dimensional supersymmetries rather than all
superconformal charges, but together with the $SO(4,2)$
isometries of the ansatz these imply the full $SU(2,2|1)$
superconformal invariance. One wonders at first whether the
$\cn=4$ 5d gauged supergravity\footnote{This denotes 32
supercharges, which in the supergravity literature is denoted as
$\cn=8$, even though the minimal 5d supergravity has 8
supercharges.}, which is often used to analyze RG flows, could be
useful here, but it turns out that it is not. The gauged
supergravity is believed to be a consistent truncation of the 10d
IIB supergravity in the $\ads$ background, and it is known how to
identify its fields at first order with the 10d modes (but not at
higher orders). Once we know our deformation at first order, which
turns out to be a 2-form potential mode on \sfv in the $({\bf
1,45})$ representation of $SO(4,2) \times SO(6)_R$, we can check
whether this mode is retained by the gauged supergravity.
However, it is easy to see that it is not, since the $SO(6)$
gauged supergravity does not have scalars in the ${\bf 45}$ (and
the more exotic $SO(4,2)$ gauging does not have a two-form in the
${\bf 1}$ either).

Next, we construct the perturbation problem by expanding the
fields (and the covariantly constant spinor) in a perturbation
parameter $h$, substituting into the equations and attempting to
solve them order by order. Two crucial ingredients are the full
form of the SUSY variation equations for type IIB found by Schwarz
\cite{SchwarzIIB}, and the list of $AdS_5$ scalars in this
background (these are our variables) found by Kim, Romans, and van
Nieuwenhuizen \cite{Kim:1985ez}. The $SO(4,2)$ isometry restricts
our solution to depend only on the \sfv coordinates, and
supersymmetry mandates an additional $U(1)_R$, which effectively
restricts the problem further to ${\bf CP}^2$ through the Hopf
fibration.

The first order analysis was performed in \cite{GranaPolchinski},
and the solutions to the linearized equations $L\, \phi^{(1)}=0$
are precisely the marginal deformations which preserve
supersymmetry ($\phi$ stands for all the fields, and the
superscript $(1)$ denotes the order in the perturbation). At
higher orders the equation is $ L\, \phi^{(k)} =\dots$, where the
right-hand side depends only on quantities of a lower order than
$k$ (possibly we have also constraint equations $0=\dots$). These
equations can be solved at each order, except for the zero modes
of $L$ and the constraint modes, and the crucial point is whether
the right-hand side vanishes for these modes. If these modes on
the right-hand side are non-zero at order $k$ then the equations
cannot be solved beyond this order, and we say that we have an
{\it obstruction}.

At first it seems like an obstruction can appear at any order so
that an exact solution would have to survive all orders, which
would be infinitely improbable. However, the correct picture is
that any given obstruction can eliminate solutions only once --
{\it ``the obstruction only shoots once''}. To understand this we
should go back and think about the whole system of non-linear
equations, rather than consider the perturbation series. The
obstruction amounts to adding a finite number of constraints
beyond the linear approximation. We know that generically
$\mbox{dim}(\cM_c)=\mbox{\#(zero modes)\ -\ \#(additional
constraints)}$. The solution of a higher order constraint
(without any linear terms) generically leaves a singularity at
the origin. Therefore, once we have an obstruction the original
perturbative series does not make sense anymore, and cannot be
used to obtain obstructions at higher orders.\footnote{As a
simple example to have in mind consider the equation
$x^2-y^2-x^3=0$. At the linear level there are two zero modes $x$
and $y$. At second order we find an obstruction given by the
singularity equation $x^2-y^2=0$, whose solution is $y=\pm x$.
Naively, at third order we get an additional obstruction, but in
fact the way to continue is to re-organize the perturbation
expansion by choosing one of the branches and to expand around
it. Then, there are no additional obstructions. For instance, we
can choose $y=x+\epsilon(x)$ and solve perturbatively for
$\epsilon$, finding $\epsilon(x)=-x^2/2 + \dots$.}

In section \ref{perturb} we turn to an explicit order by order
study of the equations. The equations, whose variables are the
bosonic supergravity fields and the covariantly constant spinor,
happen to have an important $\IZ_2$ symmetry. A given field can
appear in the perturbation expansion either at odd orders or at
even orders, and for the covariantly constant spinor this
decomposition coincides with its four dimensional chirality
decomposition (namely, the positive chirality spinor gets
contributions at even orders, and the negative chirality spinor
at odd orders). At first order the 3-form and the negative
chirality spinor get turned on. At second order we solve for the
complex string coupling\footnote{This was done already in
\cite{GranaPolchinski}.}, the warp factor, the metric
perturbation and the positive chirality spinor. The construction
of a second order conformal solution for the cases which are
supposed to correspond to exactly marginal deformations may be
viewed as a verification of the AdS/CFT correspondence.

At second order we find that a certain potential function which
contributes to the metric receives a $\log(r/r_0)$ contribution
(where $r$ is the radial coordinate of $AdS_5$) precisely when
the (quadratic) field theory equation for exact marginality is
not satisfied. However, this $\log(r)$ dependence seems to
disappear from all observables\footnote{In our computation we do
not gauge fix most of the diffeomorphism symmetries of the
problem, including diffeomorphisms which do not vanish on the
boundary of AdS space. It is possible that in a more careful
computation such diffeomorphisms should be constrained, and then
we would find a different solution, related by a diffeomorphism
to ours, which would explicitly include $\log(r)$ factors
breaking conformal invariance at second order.}. So, it seems
that if we do not impose additional constraints we find a second
order solution even when the equation is not satisfied. We
explain why this result is consistent with our field theory
expectations, and we show explicitly that at third order in the
deformation a solution is possible only if the field theory
condition for exact marginality is satisfied. This arises from a
factorization of our third order result that we do not know how
to explain directly from the supergravity point of view.

In section \ref{non-renorm} we discuss the third order solution
in the case where the exact marginality condition is not
satisfied. This solution involves a logarithmically running
coupling, and we compare it with the expected result from
perturbation theory. We find a surprising agreement between the
two, even though the field theory result is derived at weak
coupling and the supergravity result is at strong coupling. Our
results clearly show that the leading deformation-dependent
coefficient in the beta function is independent of the 't Hooft
coupling in both limits, and an explicit computation shows that
even the numerical coefficient is the same.
 This suggests that there is some non-renormalization theorem for the leading-order
term in the beta function (and in the anomalous dimension), but
we have not been able to show this directly (or to relate it to
any of the known non-renormalization theorems).

Finally, in appendix A we study the singularity at the origin of
$\cM_c(\cn=4\,)$, some relevant finite subgroups of $SU(3)$, and
$SU(3)$-invariant coordinates for $\cM_c$.

To summarize our results, we find an explicit solution for the
deformations at second order and display the appearance of the
obstruction. Our analysis, however, does not prove that no other
obstructions exist, and that the solution we find can be extended
to all orders. It would be interesting to investigate this problem
further, and to attempt to find an exact (all-orders) solution
for the case of exactly marginal deformations.

\sbsection{Related topics}

If \sfv had Einstein deformations (or Sasaki-Einstein
deformations if we include SUSY), then these would satisfy the
equations of motion of type IIB after taking the direct product
with $AdS_5$. However, it is known that \sfv is Einstein rigid --
it has no such deformations (actually non-rigid positive Einstein
manifolds are not easy to find -- none were known until the
1980's \cite{Besse}, but by now many examples are known
\cite{LeBrun}). String theory gets around this obstacle by
generalizing the Einstein condition, adding to the metric
additional fields, and incorporating them in the SUSY variation
equations.

The geometry of supersymmetric {\it solutions} of supergravity
with non-zero field strengths is not well understood yet. One
would like to generalize the familiar notions from the purely
geometric case of covariantly constant spinor, reduced holonomy,
complex structure, K\" ahler geometry and so on. In some special
cases this is understood. It is known how to add a non-trivial
vector bundle which needs to satisfy the Donaldson-Uhlenbeck-Yau
equation. Strominger \cite{StromingerTorsion} studied backgrounds
with non-zero NS 3-form field strength $H_{NS}$ and found that
they have a complex structure, but the associated K\" ahler form
$k$ is not closed, but rather $(\del-{\bar \del})k \propto
H_{NS}$. More recently, the conditions for possibly warped
``compactifications'' (over possibly non-compact manifolds) with
non-vanishing field strength for higher rank forms were found in
several cases. Note that topological constraints (on the
cohomology of the field strength) are not relevant in our case
since we are smoothly deforming a topologically trivial
background with no 3-form flux (and in any case we have no
non-trivial 3-cycles).

Another interesting feature of this problem is the change in the
amount of supersymmetry on the moduli space. For weakly coupled
string backgrounds in flat space with zero RR field strengths
this is forbidden, as can be understood either from worldsheet
arguments \cite{BanksDixon} or from a spacetime approach
\cite{DineSeiberg} which assumes only the vanishing of the
cosmological constant. Our case clearly circumvents these
assumptions. It would be interesting to understand what are the
conditions for this to happen.

\sbsection{Open questions}

The most immediate application of our work, one which we hope to
pursue, is the generalization to $AdS_4 \times S^7$. It would
also be interesting to generalize our results to orbifolds of
$\ads$ \cite{Kachru:1998ys}, which have many more exactly marginal
deformations than $\ads$, as discussed in \cite{Aharony:2002tp}.

In this paper we study the local structure of $\cM_c(\cn=4)$
around the $\cn=4$ theory -- its dimension, the first terms in
the expansion, and the proper gauge invariant coordinates. It
would be interesting to study also the geometry away from this
point, together with its global structure. Some particular points
on $\cM_c(\cn=4)$ were studied in
\cite{Berenstein:2000hy,BJL,Berenstein:2000te} and were argued to
be dual to IIB string theory on $\IZ_m\times \IZ_m$ orbifolds of
$\ads$ with discrete torsion. These orbifolds have a different
topology from $\ads$, so supergravity must break down as we go
towards these points. One can show that indeed the corrections to
supergravity become large before the solutions of
\cite{Berenstein:2000hy,BJL,Berenstein:2000te} start having a good
geometrical approximation (with $S^5 / (\IZ_m \times \IZ_m)$ much
larger than the string scale). Thus, our results are consistent
with all these backgrounds sitting together in one large moduli
space, which has different geometrical approximations in different
regions.

What is the geometry of $\cM_c$ in general?  Generally, the
scalars in vector and tensor multiplets of 5d $\cn=1$
supergravity are known to be valued in a ``very special
geometry'' manifold, while scalars in hypermultiplets are valued
in a quaternionic space. Together with the supergravity potential
these should determine the moduli space and its geometry (see
\cite{CDA} for the state-of-the-art), at least for small
deformations when supergravity is valid. Unfortunately, it does
not seem to be possible to truncate the computations to a 5d
supergravity theory with a finite number of fields, so this
analysis may be rather difficult. Alternatively, it should be
possible to derive the geometry of $\cM_c$ from field theory. The
S-duality group $\SLt{\IZ}$ is expected to identify different
points on $\cM_c$. Other interesting global issues are whether
$\cM_c$ has singularities other than the $\cn=4$ theory, and
whether it has any non-trivial topology.

Similar methods may be used to study the {\it non-supersymmetric}
deformation problem as well. Clearly $\cM_{c,\mbox{non-susy}}
\supseteq \cM_c$, and since we do not know of any symmetry to
protect additional exactly marginal directions which would break
supersymmetry it is natural to expect an equality to hold. At weak
coupling this can be shown from perturbative computations. The
way to analyze this at strong coupling would be to expand the
equations of motion of type IIB supergravity (rather than the
supersymmetry equations) perturbatively in the deformation. At
first order there are many more marginal deformations compared to
the supersymmetric case.
 In order to rule these out as exactly marginal deformations
one has to solve for the perturbation series going up to an order
where they fail. If one could show at some order (hopefully low,
 though we show that it has to be at least third order)
that all but the superconformal ones are not exactly marginal one
would be done. However, if some marginal directions are not
eliminated one would need to go to all orders to prove that they
are indeed exactly marginal, and it is hard to imagine doing that
without supersymmetry.

Another issue is the translation into field theory of our
perturbative calculation. In our computation we find that the
supergravity fields are corrected in perturbation theory in the
deformation, with apparently generic corrections (constrained by
conformal invariance and the global symmetries). These
corrections involve scalar fields coming from spherical harmonics
on the \sfv, which appear without the usual radial dependence
they would have, had they appeared at first order. It is not
clear how to interpret the computed higher order corrections in
the field theory.

\section{Field Theory Analysis}
\label{FT}

The Lagrangian for an $SU(N)$ gauge theory with $\cn=4$
supersymmetry is uniquely determined up to the choice of the
gauge coupling $g_{YM}$ and the theta angle $\theta$, which can
be joined into a complex coupling $\tau \equiv {\theta \over
2\pi} + {4\pi i \over g_{YM}^2}$. For any value of $\tau$ the
theory is exactly conformal, so this coupling is an exactly
marginal deformation. The $\cn=4$ vector multiplet includes a
vector field, four adjoint Weyl fermions (in the $\bf 4$
representation of the $SU(4)$ R-symmetry) and six real scalars
$\varphi$ (in the $\bf 6$ representation). We can write the
Lagrangian for $\cn=4$ SYM in $\cn=1$ superspace notation. The
$\cn=4$ vector multiplet splits into an $\cn=1$ vector superfield
and three chiral superfields $\Phi^1,\Phi^2,\Phi^3$ in the
adjoint representation. In addition to the usual $\cn=1$ kinetic
terms, the $\cn=4$ theory has a superpotential of the form
 \be W = {1\over 6}{\tilde h} \epsilon_{ijk} \tr(\Phi^i \Phi^j \Phi^k), \ee
with $\tilde h = g_{YM}$ (if we choose a canonical normalization
for the kinetic terms of the $\Phi^i$). When we write the theory
in $\cn=1$ language only an $SU(3)\times U(1)_R$ subgroup of the
full $SU(4)$ R-symmetry is manifest, with the $SU(3)$ rotating
the chiral superfields $\Phi^i$. Our conventions for $SU(3)$ and
$SU(4)$ representations, and the relations between them, are
summarized in appendix B.

We are interested in studying additional exactly marginal
deformations of this theory. For $N \geq 3$, the (classically)
marginal deformations which preserve SUSY include, in addition to
the gauge coupling constant $\tau$ and the superpotential
coefficient $\tilde h$ discussed above, ten coefficients
appearing in a superpotential of the form
 \be W = {1\over 3} h_{ijk} \tr(\Phi^i \Phi^j \Phi^k) \ee
with symmetric coefficients $h_{ijk}$. To check for exactly
marginal deformations we can start by analyzing which
deformations are marginal at 1-loop. The matrix $\gamma$ of
anomalous dimensions of the fields $\Phi^i$, arising from the
wave--function renormalization of these fields, is in the
$\bf{8}+\bf{1}$ representation of $SU(3)$. The
non--renormalization theorem for the superpotential implies that
(if we rescale the fields so that their kinetic terms remain
canonical) the coupling constant running is given by
\be \label{beta} \beta_{h_{ijk}} = -{1\over 2} (h_{ijl}
\gamma^l_k + h_{ilk} \gamma^l_j + h_{ljk} \gamma^l_i). \ee
This is required to vanish for exactly marginal deformations, so
the gamma function also has to vanish in this case. At 1--loop
order the traceless part of the gamma function, in the $\bf{8}$
representation of $SU(3)$, depends only on the couplings
$h_{ijk}$, so its vanishing provides 8 constraints on these
couplings, of the form (assuming that the fields are canonically
normalized)
\be \label{obstone} \gamma_i^j = -{{N^2-4} \over {64 N \pi^2}}
h_{ikl}\, \bh^{jkl} = 0, \ee
where $\bh^{ijk}$ is the complex conjugate of $h_{ijk}$, and the
equation is to be taken in the $\bf 8$ representation of $SU(3)$
(the singlet $\gamma^i_i$ in $\gamma^i_j$ includes also
contributions from the gauge coupling and from ${\tilde h}$). The
vanishing of these components of the gamma function gives 8 real
constraints (since $\gamma_i^j = (\gamma_j^i)^*$), and by an
$SU(3)$ rotation we can remove 8 additional real degrees of
freedom, and be left with a 2 complex dimensional space of
solutions. For an appropriate choice of $\tilde h$ (which can be
chosen to be real by a global symmetry transformation) all the
beta functions vanish on this space of solutions (at one-loop).
By an appropriate $SU(3)$ rotation one can write the general
solution to (\ref{obstone}) in terms of two (complex)
coefficients $h_1$ and $h_2$, appearing in the superpotential as
\be \label{suppot} W = {h_1 \over 2} \tr(\Phi^1 \Phi^2 \Phi^3 +
\Phi^1 \Phi^3 \Phi^2) + {h_2 \over 3} \tr((\Phi^1)^3 + (\Phi^2)^3
+ (\Phi^3)^3). \ee

When (\ref{obstone}) is not satisfied, we have an anomalous
dimension for the chiral superfields at second order in $h$.
However, this anomalous dimension is not a physically measurable
quantity, since it can be swallowed into a normalization of the
fields; in fact, equation (\ref{beta}) for the running coupling
constant is usually derived by absorbing the wave function
renormalization into the normalization of the fields, after which
the coupling constant becomes scale-dependent. This
scale-dependence is a physically measurable quantity, and we see
that if (\ref{obstone}) is non-zero it occurs at third order in
the deformation parameter $h$.

In fact, the two deformations described in (\ref{suppot}) are
actually exactly marginal \cite{Leigh:1995ep} (and not just
marginal at 1-loop order as we showed above), so that the $SU(N)$
$\cn=4$ SYM theory has two (complex) exactly marginal
deformations (in addition to the $\cn=4$ flat direction
corresponding to changing the coupling). This is because if we
look at the $\cn=4$ theory deformed by the superpotential
(\ref{suppot}), then supersymmetry forces the (Wilsonian) beta
functions of $h_1$ and $h_2$ (\ref{beta}), as well as those of
the gauge coupling $g_{YM}$ and the $\cn=4$ superpotential
coupling ${\tilde h}$, to be proportional to the gamma function
of the fields $\Phi^i$. The particular deformations we discuss
preserve a $G_1 \supset \IZ_3\times \IZ_3$ symmetry given by the
transformations $\Phi^1 \to \Phi^2, \Phi^2 \to \Phi^3, \Phi^3 \to
\Phi^1$ and $\Phi^1 \to \Phi^1, \Phi^2 \to \omega \Phi^2, \Phi^3
\to \omega^2 \Phi^3$, where $\omega$ is a cubic root of
unity\footnote{An additional $\IZ_3$ symmetry which multiplies
all the $\Phi^i$ by $\omega$ is also preserved by the
deformation. The precise symmetry group $G_1$
 is actually not the
direct product of the three $\IZ_3$ factors, since their
generators do not commute; rather it is an order 27 group
generated by generators $U,\, V,\, \omega$ with the relations
$U^3=V^3=\omega^3=1$ and $U\,V = \omega\, V\, U$. We provide an
additional discussion of this group in appendix A. There is also
an unbroken $U(1)_R$ symmetry, under which the scalars in the
chiral superfields have charge $2/3$. In the conformal case this
$U(1)_R$ becomes part of the superconformal algebra.}. This
symmetry forces the gamma functions of the fields $\Phi^i$ to all
be the same and it does not allow them to mix, namely $\gamma^i_j
= \gamma \delta^i_j$. Thus, we have a single constraint
$\gamma=0$ which is sufficient to ensure conformal invariance.
Since this is a single equation in four variables, we expect
generically to have a 3-complex dimensional surface where
\be \gamma(\tau,{\tilde h},h_1,h_2) = 0, \label{gamma-in-1} \ee
which corresponds to exactly marginal deformations, namely it is
a surface of fixed points of the renormalization group
flow\footnote{We are being somewhat imprecise here since it is
not clear exactly how to define the gamma function in a
gauge--invariant way beyond perturbation theory, but we do not
expect this subtlety to affect our result. Our results below can
be viewed as evidence that considerations of this type do extend
beyond perturbation theory.}.

In general we do not know what the surface of solutions to
(\ref{gamma-in-1}) looks like, except that it includes the $\cn=4$
theory $g_{YM}={\tilde h},h_1=h_2=0$. Our 1-loop analysis showed
that at weak coupling we can turn on $h_1$ and $h_2$ (and then
determine $\tilde h$ as a function of $g_{YM}$, $h_1$ and $h_2$
from (\ref{gamma-in-1})). In fact, we can also show that for any
value of $g_{YM}$ and to leading order in the deformation away
from the $\cn=4$ fixed line, the additional exactly marginal
deformations are exactly given by $h_1$ and $h_2$.  This is
because the coupling corresponding to changing $\tilde h$ to be
different from $g_{YM}$ is, in fact, not marginal on the fixed
line for non--zero coupling. The coupling corresponding to
changing $g_{YM}$ and $\tilde h$ together is an
$SU(4)_R$-singlet, which is in a chiral primary multiplet (it is
identified with the dilaton in supergravity), so its dimension is
protected to be exactly 4 on the $\cn=4$ fixed line. The
operators coupling to $h_1$ and $h_2$ are also part of chiral
multiplets, whose dimensions do not get renormalized in the
$\cn=4$ theory. However, by examining the deformation
corresponding to changing $\tilde h$ to be different from
$g_{YM}$, one can see that it is in fact a component of a
non--chiral operator in the $\bf 15$ of $SU(4)$. It has a
non-zero anomalous dimension which can be computed in
perturbation theory, and the AdS/CFT duality
\cite{Maldacena:1998re} suggests that for large $g_{YM}^2 N$ its
dimension is at least of order $(g_{YM}^2 N)^{1/4}$, since no
field which can be identified with this operator appears in the
supergravity spectrum. In the $\cn=1$ language this can be seen
from the fact that the field $\epsilon_{ijk} \tr(\Phi^i \Phi^j
\Phi^k)$ is a descendant, since for non-zero coupling $\{\bar Q,
\bar \lambda^k\} \propto [\Phi^i, \Phi^j]$ (where $\bar
\lambda^k$ are the fermionic components of the superfields
$\Phi^k$). An explicit computation of the beta function of ${\hat
h} \equiv {\tilde h}/g_{YM}-1$ shows that $\beta_{\hat h}$ must
vanish on the fixed line but $\beta_{\hat h}/{\hat h}$, which is
the anomalous dimension of the operator coupling to ${\hat h}$ on
the fixed line, is generally different from zero.

We conclude that field theory arguments imply that there are two
exactly marginal deformations away from the $\cn=4$ fixed line,
which correspond (at leading order away from the fixed line) to
the $U(1)_R\times G_1$--preserving couplings $h_1$ and $h_2$.
Note that the fact that the couplings $h_1$ and $h_2$ (if we
choose $\tilde h$ appropriately) are exactly marginal implies
that at all orders in perturbation theory (and even
non-perturbatively) the gamma function matrix vanishes if we only
turn on these couplings. This in turn means that the gamma
function matrix vanishes (for an appropriate choice of $\tilde
h$) whenever the 1-loop term (\ref{obstone}) (corresponding to
the projection of the product $h\, \bh$ of the ${\bf 10}\times
{\bf {\overline{10}}}$ representations of $SU(3)$ onto the $\bf 8$
representation) vanishes. We will denote this term by $(h
\bh)_{\bf 8}$.

Our 1-loop analysis shows that there cannot be any additional
exactly marginal deformations which preserve
supersymmetry\footnote{At least at weak coupling; it is hard to
imagine a phase transition changing the dimension of the space of
conformal theories as one moves from weak coupling to strong
coupling, but perhaps this is not completely impossible.}. In
perturbation theory it is simple to compute the beta functions of
the various couplings perturbatively, and to find that the
couplings in $h_{ijk}$ which are not exactly marginal are, in
fact, marginally irrelevant, due to their non-zero beta function
at order $h^3$. These coupling constants are not asymptotically
free, so the corresponding theories do not make sense as a full
description of the physics, but one can still use them as an
effective description in some range of energies where the
couplings are small. In such an energy range the couplings will
flow according to their perturbative beta functions, which are
proportional to the anomalous dimensions matrix $\gamma^j_i$. We
will construct below the dual of such a flow, which will enable
us to compare in section 5
 the anomalous dimensions at leading
order in $h$ at weak and strong gauge couplings.

We end our field theory discussion by noting that it is possible
to translate the gamma functions we discussed here, which are not
gauge invariant, to a gauge-invariant language. One way to do
this is just to compute the anomalous dimensions of composite
chiral operators like $\tr(\Phi^{i_1} \cdots \Phi^{i_k})$ (in a
symmetric $SU(3)$ representation); since there are no
short-distance singularities when chiral operators are brought
together, the anomalous dimension of these gauge-invariant
operators is simply the sum of the anomalous dimensions of their
components (with an obvious generalization in case the gamma
matrix is not diagonal).

Another way to phrase our results in gauge-invariant terms is to
use the relation between the gamma functions which appear in the
beta function equations and the global symmetries which are
broken by the various couplings. In the absence of a
superpotential, a classical gauge theory has a $U(1)$ global
symmetry rotating the phase of every chiral superfield in the
theory (which is enhanced to $U(k)$ if we have $k$ chiral fields
in the same complex representation of the gauge group). Every
superpotential coupling breaks some combination of these $U(1)$
symmetries, and the gauge coupling also breaks a combination of
$U(1)$ symmetries through the axial anomaly. A superpotential
coupling $W = \lambda \prod_i (\Phi_i)^{p_i}$ breaks the linear
combination of $U(1)$'s given by $\Phi_i \to e^{i \alpha p_i}
\Phi_i$ (while preserving all orthogonal combinations), while the
gauge coupling breaks (through the axial anomaly) the linear
combination of $U(1)$'s given by $\Phi_i \to e^{i \alpha C(r_i)}
\Phi_i$, where $C(r_i)$ is the quadratic Casimir of the
representation $r_i$ of the gauge group which the field $\Phi_i$
is in.

In both cases (assuming that the 1-loop gauge coupling beta
function vanishes), if the coupling breaks the linear combination
given by $\Phi_i \to e^{i \alpha C_i} \Phi_i$, its beta function
is proportional to $\sum_i C_i \gamma_i$ (with obvious
generalizations to the case of non-Abelian global symmetries).
This follows from the superpotential non-renormalization theorem
or from the NSVZ formula. If the symmetries broken by some
couplings are not independent, their beta functions will be
linearly dependent, and this is the essence of the arguments of
\cite{Leigh:1995ep} for exact marginality. If the gamma function
$\gamma_i$ is non-zero, it means that under RG flow the field
$\Phi_i$ changes as $\Phi_i \to Z_i \Phi_i$ with $\gamma_i = \del
\log(Z_i) / \del \log(\mu)$ (for a renormalization scale $\mu$).
This rescaling of $\Phi_i$ is simply a complexified version of
the global $U(1)$ symmetry acting on $\Phi_i$. Thus, having
non-zero gamma functions in an $\cn=1$ gauge theory is the same
as having a complexified global symmetry transformation acting on
the chiral superfields in the theory. The physically meaningful
gamma functions are those appearing in the beta functions, and
these involve complexified global symmetry transformations by the
global symmetry generators which are broken by the couplings in
the theory, with coefficients whose dependence on the
renormalization scale is given by the gamma function.

In the $\cn=4$ case described above, the relevant symmetries are
the $SU(3)$ symmetry which is broken by the couplings $h$, and the
$U(1)$ symmetry rotating all the chiral superfields together
which is broken by $h$ as well as by the gauge coupling and the
anti-symmetric superpotential coupling. Non-zero gamma functions
will induce complexified $U(3)$ global symmetry transformations
acting on the chiral superfields $\Phi^i$, and it is easy to
express those in terms of their action on gauge-invariant
variables (since we know their global symmetry transformation).
The scale-dependence of the various superpotential couplings is
simply given by their transformation under these complexified
global symmetry transformations. The scale-dependence of the
gauge coupling is also proportional to a factor coming from this
transformation, which is the numerator of the NSVZ formula for
the beta function.

\section{Supergravity Analysis -- Methods}
\label{methods}

A deformation of a conformal theory which has a dual description
in terms of string theory on $AdS_5$, such as $\cn=4$ SYM theory
which is dual to type IIB string theory on $AdS_5\times S^5$,
corresponds (for large $N, g_{YM}^2 N$ where supergravity is a
good approximation to string theory) to a solution of supergravity
with appropriate boundary values for the fields
\cite{Gubser:1998bc,Witten:1998qj}. Marginal operators correspond
to massless fields in supergravity, and for marginal deformations
the corresponding massless SUGRA fields should approach a
constant at the boundary of the $AdS_5$ space, while for relevant
deformations they should decay like an appropriate power of the
radial coordinate. Having an exactly marginal deformation means
that none of the fields in the deformed solution should depend on
the AdS coordinates, since the $SO(4,2)$ symmetry should remain
exact. Thus, the fields can only obtain an angular dependence,
and are given by some combination of $S^5$ spherical harmonics.
Ideally, we would like to find exact solutions corresponding to
the exactly marginal deformations described above, but we have
not been able to do this so far. An alternative option, which we
will pursue in this paper, is to try to construct the solutions
in a perturbation series in the deformation around the $\cn=4$
fixed line.  This approach only makes sense for exactly marginal
deformations, since relevant or irrelevant deformations would not
remain small throughout the $AdS_5$ space and the perturbation
expansion would break down.

Our goal in this paper will be to show that the exactly marginal
deformations which we analyzed in field theory in the previous
section exist also in the AdS dual, in the supergravity limit. We
begin in this section by examining how to analyze marginal
deformations perturbatively in supergravity, and how to see if
they are exactly marginal or not. In the first subsection we will
analyze how to solve the supergravity equations of motion
perturbatively in the deformation, and in the second subsection
we will analyze how to solve the supersymmetry equations.

\subsection{Perturbative expansion of the equations of motion}

To leading order in the deformation away from the $\cn=4$ theory,
marginal deformations of this theory correspond to giving vacuum
expectation values (VEVs) to the massless scalars of the SUGRA
spectrum analyzed in \cite{Kim:1985ez}. The light scalars in the
supergravity spectrum are depicted below in figure 1 (which is
reproduced with permission from \cite{Kim:1985ez}). There is one
complex marginal deformation corresponding to the zero mode (on
the $S^5$) of the dilaton and the axion, which is obviously
exactly marginal also in SUGRA, since the supergravity equations
only depend on the derivatives of the dilaton (thus, this is in
fact true for any supergravity solution in the supergravity
approximation). This deformation corresponds to changing the gauge
coupling $\tau$ in the field theory. There are complex marginal
deformations in the $\bf{45}$ representation of $SU(4)_R$
corresponding to turning on a 2-form field (any combination of
the NS-NS and R-R 2-form fields) on the $S^5$ proportional to the
spherical harmonic $Y_{ab}^{[2,-]}$ (see \cite{Kim:1985ez} for
definitions of the relevant spherical harmonics), and real
marginal deformations in the $\bf{105}$ representation
corresponding to changing a particular combination of the trace
$h^a_a$ of the metric on the $S^5$, the trace $h^{\mu}_{\mu}$ of
the $AdS_5$ metric, and the 5-form field, by an amount
proportional to the spherical harmonic $Y^{[4]}$. Using the
identification between the field theory operators and the
supergravity fields \cite{Witten:1998qj} we find that the
marginal deformations $h_{ijk}$ described above involve fields in
the $\bf{45}$, to leading order in the deformation; at higher
orders there will be a mixing of the various operators. In the
$\cn=4$ field theory the corresponding operators are descendants
of the chiral primary operator $\tr( \varphi^{\{I} \varphi^J
\varphi^{K\}})$ (where the $\varphi$'s are the scalar fields in
the $\cn=4$ vector multiplet) constructed by acting on it with
two SUSY generators; this gives rise to both Yukawa couplings and
scalar potential terms (arising from the superpotential
contributions to the SUSY variations).

\EPSFIGURE{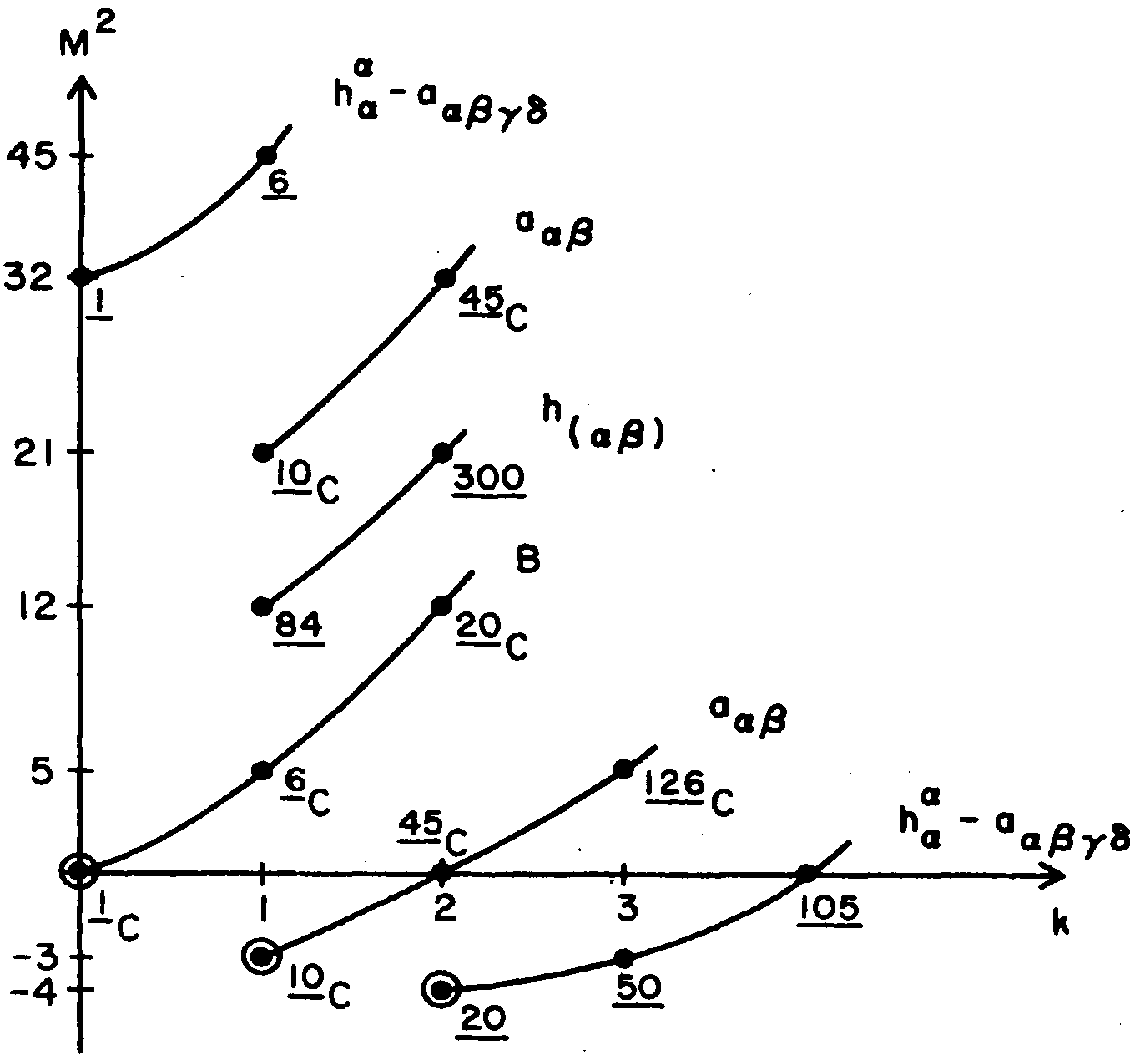}{ Spectrum of $AdS$ scalars for IIB
supergravity on $AdS_5 \times S^5$ \cite{Kim:1985ez}.
\label{KRvNfig}}

There is a well-defined procedure for constructing a SUGRA
solution, corresponding to an exactly marginal deformation,
perturbatively in the deformation parameter. Obviously, such a
deformation involves only fields which are scalars on the $AdS_5$.
In \cite{Kim:1985ez}, the SUGRA equations of motion were expanded
to first order around the $AdS_5\times S^5$ solution. The
deviation of each of the $AdS_5$-scalar fields from its
expectation value in the $AdS_5\times S^5$ solution can be
expanded in spherical harmonics, leading to a series of fields
$\phi_{j,[k]}$, where $[k]$ labels the spherical harmonic and $j$
labels the field. For each of these fields the linearized SUGRA
equations of motion then take the form \be \label{linearized}
(\bigtriangleup_{AdS} + m^2_{j,[k]}) \phi_{j,[k]} = 0. \ee The
fields $\phi_j$ include the dilaton $B$ ($B = (1 + i\tau)/(1 -
i\tau)$), the NS-NS and R-R two-form fields $A_{ab}^{1,2}$
(indices $a,b,c,\cdots$ will be taken to run over the $S^5$
coordinates), the metric components $h_{ab}$, the trace of the
metric $h^a_a$ and the 4-form field $A_{abcd}$; the last two
fields actually mix together, so the fields $\phi_j$ are in this
case appropriate linear combinations of the two \cite{Kim:1985ez}.
The spherical harmonics appearing in the expansion depend on the
tensor type (with respect to $S^5$) of the field; for $B,
h^{a}_{a}$ and $A_{abcd}$ they are the scalar spherical
harmonics, for $A_{ab}$ they are the anti-symmetric tensor
spherical harmonics, and for $h_{ab}$ they are symmetric tensor
spherical harmonics. As described in \cite{Kim:1985ez}, some of
the fields in the spherical harmonic expansion may be gauged away
by the symmetries of the SUGRA theory. In particular, the
expansion of the tensor fields could include also derivatives of
vector and scalar spherical harmonics, but these may be gauged
away. The fields in the doubleton representation (corresponding
to a free $U(1)$ vector multiplet) may also be gauged away in the
bulk of $AdS_5$. After these gauge fixings, the complete list of
fields appearing in (\ref{linearized}) is given in table III and
figure 2 of \cite{Kim:1985ez}, which we reproduced as figure 1
above.

To go beyond the linearized approximation, we need to expand the
supergravity equations of motion to higher order in the fields
$\phi_{j,[k]}$. We can do this systematically by expanding each
of these fields in a power series in a deformation parameter $h$
(which will be simply the superpotential coefficient discussed
above), of the form $\phi_{j,[k]} = \sum_{i=1}^\infty h^i
\phi_{j,[k]}^{(i)}$ (the generalization to having several
expansion parameters is straightforward). Plugging this into the
exact SUGRA equations of motion, we can expand them in a power
series in $h$, and solve them order by order.

Since we are starting with an exact solution, the equations are
satisfied to zeroth order in $h$. To first order in $h$, the
equations are exactly the linearized equations (\ref{linearized})
for the leading $\phi_{j,[k]}^{(1)}$ term. As expected, this
means that if we want no $AdS$-dependence, we must turn on only
massless fields at this order.

At the $n$'th order in this expansion, each equation will involve
a term including $\phi_{j,[k]}^{(n)}$ with no other deformation
fields, and other terms including lower order fields. The term
involving $\phi_{j,[k]}^{(n)}$ will be exactly the linearized
term in this field; thus, we will find equations of the form \be
\label{unlinearized} (\bigtriangleup_{AdS} + m^2_{j,[k]})
\phi_{j,[k]}^{(n)} = f(\{\phi_{l,[m]}^{(1)}, \phi_{l,[m]}^{(2)},
\cdots, \phi_{l,[m]}^{(n-1)} \}), \ee for some function $f$
involving various lower order fields on the right hand side. In
principle, other spherical harmonics might appear in the
expansion that have been gauged away at the leading order, so
they do not correspond to fields $\phi_{j,[k]}$; however, we can
change the gauge condition order by order to take account of such
terms, and concentrate only on the equations of the form
(\ref{unlinearized}). There is a subtlety here, which is that in
order to do this we may need to perform gauge transformations
that do not vanish at infinity, and it is not clear if these
should be allowed or not.
 If we do not allow these gauge
transformations, we might find an obstruction from the ``gauge
modes'' to the construction of a solution. Here we assume that
all gauge transformations are allowed.

Since we are looking for $AdS$-independent solutions, only the
mass term contributes on the left hand side, so for massive
fields the equation simply determines the form of
$\phi_{j,[k]}^{(n)}$. However, for massless fields, if the
expression on the right hand side of the equation is non-zero,
there is no $AdS$-independent solution. Thus, at each order in
the perturbative expansion, we get one ``obstruction'' equation
which the deformations must satisfy for each massless field in
the SUGRA theory\footnote{The ``obstruction'' can be thought of
as a single non--linear equation which has to be satisfied, and
at each order in perturbation theory we encounter the $n$'th
order terms in this equation.}. To check if a deformation is
exactly marginal or not we should plug in the appropriate first
order deformation, solve the equations order by order, and see if
we encounter obstructions from the higher order massless field
equations or not. If we find a non-zero term on the right-hand
side for a massless field, we can still solve the equations but
the fields will necessarily acquire a dependence on the radial
coordinate, breaking the conformal invariance; this corresponds
to cases with dimensional transmutation in the field theory.

As an example of this procedure let us assume that we are turning
on a generic marginal deformation of $AdS_5\times S^5$, and let us
look at the simplest equation of motion of type IIB supergravity,
\be \label{dilaton} D^M P_M = G_{ABC} G^{ABC}, \ee where $P_M =
D_M B / (1 - |B|^2)$ and $G$ is some combination of the 3-form
field strengths (we will use the conventions of
\cite{Green:1987mn} for type IIB supergravity). In the linear
approximation, this equation is simply $(\bigtriangleup_{AdS} +
m^2_{[k]}) B_{[k]}^{(1)} = 0$, where $B_{[k]}$ is the mode of the
dilaton in the $k$'th spherical harmonic on the sphere, and
$m^2_{[k]} \propto k(k+4)$, so we can only turn on the constant
$k=0$ mode. At second order, the equation will only depend on the
fields in the $\bf{45}$ representation (recall that to leading
order we can only turn on massless fields), so it will be of the
form \be \label{ndilaton} (\bigtriangleup_{AdS} + m^2_{[k]})
B_{[k]}^{(2)} = (G_{abc}^{(1)} (G^{abc})^{(1)})_{[k]}, \ee where
on the right hand side we have a projection of $G^2$ on the
appropriate spherical harmonic. Plugging in the explicit form of
the deformation in the $\bf{45}$, we find that the right hand
side is a combination of second order ($x^{\{I} x^{J\}}$) and
fourth order ($x^{\{I} x^J x^K x^{L\}}$) scalar spherical
harmonics, with coefficients that are the Clebsch-Gordan
coefficients for the $\bf 20^\prime$ and $\bf 105$
representations (respectively) in the product $\bf{45}\times
\bf{45}$. Thus, in this case we find no obstruction to solving
the equations, since the singlet spherical harmonic
(corresponding to the massless field $B_{[0]}$) did not appear on
the right hand side. However, we find that generically we will
need to turn on $B_{[2]}^{(2)}$ and $B_{[4]}^{(2)}$, which will
be given by some quadratic function of the deformation
parameters. It is not clear if turning on these constant (on
$AdS_5$) values for some modes of the dilaton has any
interpretation in the field theory -- turning on the fields
$B_{[2]}$ and $B_{[4]}$ with a particular radial dependence
corresponds to deformations or VEVs in the $\cn=4$ theory, but
the interpretation of the constant mode of these fields is not
clear. Presumably, it is related to operator mixing which occurs
after the deformation. In section 4 we will compute such
corrections to various fields in the case of the supersymmetric
exactly marginal deformations, and it may be interesting to find
a field theory interpretation for the results.

Similarly, we may analyze the other SUGRA equations. For the most
general possible marginal deformation that we could turn on, at
second order in the deformation we find no obstruction from the
equations for the fields in the $\bf{1}$ and the $\bf{45}$ (the
latter fact follows from simple group theory arguments), but
there is a possible obstruction from the equation for the
massless field in the $\bf{105}$. The coefficient of this
obstruction involves the Clebsch-Gordan coefficients for
$\bf{105}\times \bf{105} \to \bf{105}$; this gives $105$
quadratic equations in the coefficients of the first order
deformation in the $\bf{105}$ that must be satisfied for the
deformation to be exactly marginal. At third order, all the
massless field equations are in principle non-trivial, so we find
$105+45*2+1*2=197$ trilinear equations in the deformation
parameters which must be satisfied, and so on at higher orders.
Unfortunately, the resulting equations seem very complicated, so
we cannot write them down explicitly.

The equations simplify to some extent if the deformation we turn
on is only in the $\bf{45}$ representation; recall that (for some
particular elements of the $\bf{45}$) this is the deformation
that is expected to be exactly marginal from the field theory
arguments\footnote{Actually, the exactly marginal deformation
which we analyzed in section 2 is supersymmetric, which means
that it corresponds not just to an operator in the $\bf{45}$ but
one also has to add in quadratic order in the deformation an
operator in the $\bf{105}$, corresponding to the scalar potential
in the field theory. The coefficient of this operator is not
uniquely determined by the deformation in the $\bf{45}$, but
rather it depends on which $\cn=1$ supersymmetry we wish to
preserve (this is clear since the scalar potential is
proportional to $h\, \bh$, but the product of ${\bf{45}}\times
{\bf{\overline{45}}}$ does not contain the $\bf{105}$
representation). Since this additional deformation is at second
order in $h$, this does not change the qualitative discussion of
this section. A similar situation occurs with the supersymmetric
mass term, which includes not only an operator in the $\bf{10}$
but also an operator in the $\bf{20'}$ representation
\cite{Aharony:1999ti}.}. In this case, as described above, the
second order equations may always be satisfied\footnote{Up to
possible obstructions coming from the gauge-fixing conditions.},
and they generically lead to the generation of $B^{(2)}$ in the
$\bf{20^\prime},\bf{105}$ representations, of $h_{ab}^{(2)}$ in
the $\bf{84},\bf{729}$ representations and of
$(h_{a}^{a})^{(2)},A_{abcd}^{(2)}$ in the $\bf{1},\bf{20^\prime}$
representations.  These are the only representations that may
appear from group theory arguments; apriori it is not clear if
their coefficients in the actual SUGRA equations are all
non-zero, but this seems likely since there is no reason for them
to vanish. $A_{ab}^{(2)}$ is zero in this case (this is because
the equations have a $\IZ_2$ symmetry inverting the sign of
$A_{ab}$). The first constraints on deformations in the $\bf{45}$
come from the equation for the massless field in $A_{ab}^{(3)}$;
they involve the projection onto the $\bf{45}$ representation
(spherical harmonic) of the product of the fields generated at
the second order with the first order field in the $\bf{45}$ (or
its conjugate in the $\bf{\overline {45}}$; the deformation we are
turning on is of course hermitian, so it involves both fields).
This leads to 45 trilinear equations in the coefficients of the
deformation; apriori there is no obvious reason why these
equations should have non-zero solutions, but it is not impossible
from the supergravity point of view (and our field theory
analysis of the previous section implies that it should indeed
happen). At the fourth order we get 105 quartic equations from
the equations for $(h_a^a)^{(4)}, A_{abcd}^{(4)}$, and so on.
There is no obvious reason from the SUGRA point of view why all
these equations should have any non-trivial solutions.

In section 4 we will solve some of these equations to third order
and see that for specific choices of the deformation
(corresponding to the exactly marginal deformations in the field
theory) there are non-trivial solutions. However, first we will
simplify the equations by using the fact (which we have not used
so far) that we are interested in deformations that preserve some
supersymmetry.

\subsection{Perturbative expansion of the supersymmetry equations}

In our analysis above we saw that general marginal deformations
are in the $\bf{1}$, $\bf{45}$ and $\bf{105}$ representations of
$SU(4)$, and each one leads to an obstruction as well (at every
order in the perturbation expansion). For the particular marginal
deformations discussed in section 2 which preserve supersymmetry,
a $U(1)_R$ subgroup of $SU(4)$ (which is part of the
superconformal algebra) is also preserved, so all the scalars
which are charged under $U(1)_R$ will automatically be zero, and
only the fields neutral under $U(1)_R$ participate. That limits
the above set to $\bf{1}$, $\bf{8+10}$ and $\bf{27}$
representations of the global $SU(3)$ group, arising from the
$\bf{1},\bf{45}$ and $\bf{105}$ representations of $SU(4)$,
respectively. When we use later the full supersymmetry equations
in addition to the $U(1)_R$ invariance, we will find that the only
supersymmetric deformation (in addition to the string coupling in
the $\bf{1}$) is the $\bf{10}$ mentioned in section \ref{FT}.

As discussed above, from the field theory we would expect that
the only obstruction should be in the $\bf{8}$ representation, of
the form \be [(h \bh)_{\bf 8}]_i^j \equiv h_{ikl}\, \bh^{jkl} -
{1\over 3} (h_{klm}\, \bh^{klm})\, \delta_i^j = 0, \ee
 since there should be a superconformal theory whenever this is satisfied.
For the consistency of the AdS/CFT correspondence it must be the
case that all the obstructions in the supergravity actually
vanish when the quadratic expression $(h \bh)_{\bf 8}$ vanishes,
and we expect that this is the only case when all these
obstructions vanish, but we do not know how to show this directly.

To make further progress we will use the additional information
we have, which is that the deformation also preserves $\cn=1$
superconformal symmetry. In addition to the bosonic $SO(4,2)$ and
$U(1)_R$ symmetries utilized above, this includes also fermionic
generators, and we should be able to find a corresponding
generalized Killing spinor $\epsilon$.
 Again, we can analyze this
in a perturbation expansion in $h$, and in an expansion in
spherical harmonics. The expansion of $\epsilon$ will involve
spinor spherical harmonics, which are described (with their
relation to the bosonic spherical harmonics) in \cite{Kim:1985ez}.

The equations we need to solve are the vanishing of the SUSY
variation of the dilatino $\Llambda$ and the gravitino $\Psi_M$
\cite{SchwarzIIB},
\bea \label{susy} \delta \Llambda &=& {i\over \kappa} \Gamma^M
P_M \epsilon^* - {i\over {24}} \gsl \epsilon, \non \delta \Psi_M
&=& {1\over \kappa}(D_M -{i\over 2}Q_M)\epsilon + {i\over
{480}}\fsl \Gamma_M \epsilon - {1\over {96}}(2\gsl
\Gamma_M+\Gamma_M \gsl)\epsilon^*, \eea
where $\gsl \equiv \Gamma^{MNP} G_{MNP}$, $\fsl \equiv
\Gamma^{M_1 M_2 M_3 M_4 M_5} F_{M_1 M_2 M_3 M_4 M_5}$, and
$\epsilon$ is a Weyl spinor, obeying $\Gamma_{11}\epsilon =
-\epsilon$ ($\Llambda$ and $\Psi_M$ are also Weyl spinors with
$\Gamma_{11} \Llambda = \Llambda$, $\Gamma_{11} \Psi_M =
-\Psi_M$). The notations of this equation are described in detail
in section \ref{eqsvars} below. We are looking for solutions
which preserve superconformal invariance. Note that any solution
to (\ref{susy}) is also a solution to the supergravity equations
of motion, so solving these equations is enough for our purposes.

Since we are looking for a superconformally invariant solution it
is convenient to decompose the spinor in a way which takes this
into account. One way to do this, described in \cite{Kim:1985ez},
is to partition the $10$ directions into $5+5$, and write the 10
dimensional gamma matrices in terms of two groups of 5
dimensional gamma matrices. This decomposition does not seem to
be very convenient for our purposes, and instead we will use here
a different decomposition, as $10=4+6$, writing the gamma
matrices as the outer product of $SO(3,1)$ gamma matrices (which
are $4\times 4$) and $SO(6)$ gamma matrices (which are $8\times
8$). This is the same decomposition appearing in
\cite{GranaPolchinski} and it readily allows the description of
flows when conformal invariance is violated. The 10 dimensional
chirality matrix $\Gamma_{11}$ may be written as a product
$\Gamma^{11} = -\Gamma^4 \otimes \Gamma^7$ of four dimensional
and six dimensional chirality matrices, so that the spinor
$\epsilon$ will have components with $\Gamma^4$ and $\Gamma^7$
eigenvalues both positive or both negative.

In the $AdS_5\times S^5$ background, the zeroth order dilatino
equation is trivially solved, and the zeroth order gravitino
variation equation (we will write the $AdS_5\times S^5$
background as $ds^2 = Z^{-1/2} \eta_{\mu \nu} dx^\mu dx^\nu +
Z^{1/2} dx^m dx^m$ where $\mu,\nu=0,\cdots,3$, $m,n=4,\cdots,9$,
$Z = R_0^4 / r^4$ and $r^2 = x^m x^m$) says that
\cite{GranaPolchinski}
\bea \label{gravvar} 0 = \kappa \delta \Psi_\mu &=& \del_\mu
\epsilon + {1\over 2} \Gamma_\mu \Gamma^m {\del_m r \over r} (1 -
\Gamma^4) \epsilon, \non 0 = \kappa \delta \Psi_m &=& \del_m
\epsilon - {1\over 2} \epsilon {\del_m r \over r} + {1\over 2}
\Gamma^n {\del_n r \over r} \Gamma_m (1 - \Gamma^4) \epsilon. \eea
The solution to this is \be \label{susysol} \epsilon =
(R_0/r)^{-1/2} \zeta \otimes \chi, \ee where $\zeta$ is a
constant $SO(3,1)$ spinor with positive $\Gamma^4$ chirality, and
$\chi$ is a constant $SO(6)$ spinor with positive $\Gamma^7$
chirality. There are four possible constant $SO(6)$ spinors so we
get four $d=4$ supersymmetries. After we deform only one of these
will be preserved, so we choose the zeroth order SUSY parameter
$\epsilon$ to be of the form (\ref{susysol}) where $\chi=\chi_0$
is a particular one of the four possible constant $SO(6)$ spinors.

The bosonic field equations have an $SL(2,\IR)$ symmetry, with a
$\IZ_2$ subgroup that acts by changing the sign of the 3-form
field $G$. Thus, it is clear that when we turn on a first order
deformation involving purely the 3-form field $G$, as we are
doing in the case of the SUSY marginal deformations, then in the
perturbation expansion $G$ will only get contributions in odd
orders in the deformation parameter $h$, while the other SUGRA
fields (the metric, dilaton and 5-form) will only get
contributions at even orders in $h$. We can choose the matrix
$\Gamma^4$ to be purely imaginary, so that $\epsilon^*$ has an
opposite $\Gamma^4$ eigenvalue from $\epsilon$. The vanishing of
(\ref{susy}) then implies that at even orders in the deformation
$\epsilon$ has a positive $\Gamma^4$ eigenvalue, while at odd
orders $\epsilon$ has a negative $\Gamma^4$ eigenvalue (and
$\epsilon^*$ has a positive $\Gamma^4$ eigenvalue)
\cite{GranaPolchinski}.

As in the previous subsection, we can expand the SUSY variation
equations (\ref{susy}) in a power series in $h$, and at higher
orders we may encounter obstructions. In the SUSY variation
equations
the obstructions are fermionic. We will linearize the
equations (\ref{susy}) in full generality in section 4.4 below,
but for now let us discuss a specific type of obstruction that
could occur (additional obstructions will be discussed in the
next section).
Let us analyze the second equation of (\ref{susy}) in the case
that $M$ is an $S^5$ coordinate. Suppose that we knew all the
supergravity fields up to order $n$ in the deformation parameter
$h$, and that we know $\epsilon$ up to order $n-1$. Then, this
equation gives an equation for $\epsilon$ at $n$'th order,
analogous to the equation (\ref{unlinearized}) that we analyzed
above, of the schematic form
\be \label{epsunlin} (\del_a + (A_{[k]})_a^b \Gamma_b)
\epsilon_{[k]}^{(n)} = g_a(\{
\phi_{l,[m]}^{(1)},\cdots,\phi_{l,[m]}^{(n)},\epsilon^{(0)},\cdots,
\epsilon^{(n-1)} \}), \ee for some matrix $A_{[k]}$. For most
spherical harmonics this will just give a linear equation
determining $\epsilon_{[k]}^{(n)}$. As above, this will not
happen only for the case when the differential operator $(\del_a
+ (A_{[k]})_a^b \Gamma_b)$ has zero eigenvalues, and then we will
get an ``obstruction'' equation for the vanishing of the
corresponding spherical harmonic on the right--hand side, which
must be satisfied (as we found for the massless fields in the
previous subsection). By studying the case of $n=0$ it is clear
that one case in which this will happen is for $\epsilon$ in the
$\bf 4$ representation of $SU(4)$, since this is the only
solution to the spinor equations in the original $AdS_5\times
S^5$ theory. Thus, at each order in perturbation theory we get a
possible obstruction from this equation in the $\bf 4$ of $SU(4)$
(actually this obstruction will only appear at even orders). In
fact, for the deformations we are interested in which preserve a
$U(1)_R$ in $SU(4)$, the obstruction can only be in the singlet
appearing in the decomposition ${\bf 4} \to {\bf 1} + {\bf 3}$.
Thus, for the case we are interested in, the second equation with
an $S^5$ index gives a singlet obstruction at every even order of
perturbation theory, and otherwise determines $\epsilon^{(n)}$
given the other fields at order $n$. We will see below that there
are additional obstructions at odd orders.

In this paper we will not analyze the general structure of the
obstructions, but rather we will just expand the equations and
identify any obstructions when they show up.

\section{Supergravity - Perturbative Computations}
\label{perturb}

Our objective is to find deformations of type IIB supergravity on
$AdS_5 \times S^5$ which preserve the $SO(4,2)$ isometries of
$AdS_5$ and 8 real supercharges (out of 32).

In section \ref{FT} we saw that $\cn =4$ $SU(N)$ SYM theory (with
$N \geq 3$) has $10$ complex\footnote{$11$ at zero coupling,
where the anti-symmetric superpotential is also marginal.}
super-marginal operators (marginal operators preserving $\cn=1$
supersymmetry) besides the complex gauge coupling. Two out of
these are exactly marginal, given by \be  \label{conformal} (h
\bh)_{\bf 8}=0, \ee while the others fail at third order to be
exactly marginal (they are marginally irrelevant). In section
\ref{methods} we discussed some general features of the
perturbative expansion in supergravity.

In the current section we present this computation in full
detail, and find the following results :
\begin{itemize}
\item The full supergravity solution at second order, which
includes the metric, the warp factor, the dilaton (which was
already found in \cite{GranaPolchinski}) and the Killing spinor.
\item A specific obstruction is shown to eliminate the non-exactly-marginal
deformations, and we compute the third order beta function
arising in this case.
\end{itemize}
However, so far we were not able to prove that there is no
obstruction to extending the solutions obeying (\ref{conformal})
to all orders.

\subsection{Equations and variables}
\label{eqsvars}

We are looking for backgrounds preserving SUSY, and hence our
basic equations are the vanishing of the supersymmetry variation
of the fermionic fields in type IIB supergravity. These, together
with an ansatz with $SO(4,2)$ isometry, will guarantee that our
solutions will solve the type IIB supergravity equations of
motion and will have the full superconformal symmetry. The
equations are of the form \cite{SchwarzIIB} :
 \bea\label{susyvar}
 \delta \Psi_M &=& {1 \over \kappa} (\hat{D}_M -{i \over 2}Q_M)\epsilon +
{i \over 480} \hat{F}\Slash \,\Gamma_M \,\epsilon - {1 \over 96}
( 2 \hat{G} \Slash \,\Gamma_M +\Gamma_M \,\hat{G} \Slash )
\epsilon^* + \mbox{fermions}, \non \delta \Llambda &=& {i \over
\kappa} \hat{P}\Slash \,\epsilon^* - {i \over 24} \hat{G}\Slash
\,\epsilon.
 \eea
The hatted quantities $\hat{G},\hat{F},\hat{P}, \hat{D}$ denote
supercovariant versions of the unhatted quantities which differ
from them only by fermionic terms. After setting the fermions to
zero and setting $\kappa=1$ (GR units) we get \bea
 \delta \Psi_M &=& (D_M -{i \over 2}Q_M)\epsilon + {i \over
480} F\Slash \,\Gamma_M \,\epsilon - {1 \over 96} ( 2 G \Slash \,
\Gamma_M + \Gamma_M \,G \Slash) \epsilon^*, \non
 \delta \Llambda &=& i P \Slash \,\epsilon^* - {i \over 24} G\Slash \,\epsilon,
 \eea
where \bea
 D_M &=& \partial_M +{1 \over 4} \omega_M ^{NP} \: \Gamma_{NP},  \non
 G &=& (F_3 - B \,F_3^*)/\sqrt{1-|B|^2}, ~~~ F_3=dA_2,  \non
 G \Slash &=& \Gamma^{PQR} \, G_{PQR},  \non
 F &=& dA_4 - {1 \over 8} \mbox{Im}(A_2 \wedge F^*_3), ~~~~~ F=+*F, \non
 F \Slash &=& \Gamma^{MNPQR} F_{MNPQR}, \non
 Q_M &=& \mbox{Im}(B \,\partial_M  B^*)/(1-|B|^2),  \non
 P_M &=& \partial_M B/ (1-|B|^2).
\eea

Our notations are : $\omega$ is the spin connection; $B$ is the
complex scalar related to the complexified string coupling $\tau
\equiv {i\over g_s}+\chi$ by a conformal mapping of the upper
half plane to the unit disk, $B=(1+i\tau)/(1-i\tau)$;
 $G$ (or $F_3$) is the complex 3-form field strength related to the
complex 2-form potential $A_2$ which combines the NS-NS and the
RR two forms;
 $F$ is the self-dual 5-form field strength; $\Psi_M$ is the complex
 gravitino; $\Llambda$ is the complex dilatino; and $\epsilon$ is
 the complex Killing spinor (including two Majorana spinors).

The Clifford algebra is defined as usual by $\{ \Gamma_M,\Gamma_N
\}=2g_{MN}$. The $\Gamma_{11}$ chiralities of the spinors are
$+,-,-$ for $\Llambda,\Psi_M,\epsilon$ respectively
($\Gamma_{11}=\Gamma^0 \dots \Gamma^9, ~\Gamma_{11}^2=+1$), and
the signature of the metric is $(+,-,\dots,-)$. Products
$\Gamma^{M_1 \cdots M_n}$ of gamma matrices are defined to be
anti-symmetrized and divided by the order of the permutation
group, such that when all indices are different $\Gamma^{M_1
\cdots M_n} = \Gamma^{M_1} \cdots \Gamma^{M_n}$.
 These equations are written in a Majorana basis. In
order to write them in any other basis we need to replace
$\epsilon^* \to C \epsilon^*$ where $C$ is the charge conjugation
matrix in that basis.

\subsection{The ansatz}

\noindent {\it The metric}

The $AdS_5 \times S^5$ solution is given by \bea
 ds^2 &=& R_0^{~2} \, ds^2_{AdS_5} - R_0^{~2} \, ds^2_{S^5}=
{r^2 \over R_0^{~2}} dx^\mu dx_\mu - {R_0^{~2} \over r^2} dx^\al
dx_\al, \non
 F &=& \pm {1 \over R_0} ((r/R_0)^3\, dx^0\, dx^1\,
dx^2\, dx^3\, dr + (R_0/r)^5\, dx^4 \dots dx^9 \cdot dr), \non
 B &=& B_0, \label{AdS5S5} \eea
with all other fields vanishing.
 In this familiar form the $AdS_5$ is composed out of $\IR^4$
together with the radial coordinate of $\IR^6$, and hence we
refer to it as the 4-6 split (compared to a possible notation
with a 5-5 split). The indices $\mu,\nu,\dots$ run over
$0,\dots,3$, while $\al,\bt,\dots$ run over $4,\dots,9$, $M,N$
run over the whole range $0,\dots,9$ \footnote{Later we will
introduce also lower case Roman indices $i,j,\dots$ to denote
holomorphic indices in $\IR^6$.} and $r^2=x_\al x^\al$ is the
radial coordinate in $\IR^6$. The free $\pm$ sign allows to choose
between the D3 and anti-D3 solution (later we choose $-$). The
$\cdot$ stands for contraction of the 1-form $dr$ with the 6-form
on its left through the flat $\IR^6$  metric, namely
$F_{0123\al}=-r^2\, x^\al /R_0^4, ~F_{\al_1 \dots \al_5}=-
\epsilon_{\al_1 \dots \al_6} \,x^{\al_6} R_0^4/r^6$. This
solution contains one free dimensionful parameter (as always in
classical GR after setting $G=c=1$) which is $R_0$, the size of
$AdS$ and the sphere. From now on we will set $R_0=1$, and it can
always be restored by dimensional analysis (we will restore the
factors of $R_0$ in the next section when comparing to field
theory). The complex coupling $B$ can have any constant value,
and in supergravity all of them are related by the $SL(2,\IR)$
symmetry (this symmetry is broken down to $SL(2,\IZ)$ in string
theory). We will use this freedom to set $B_0=0$ (or $\tau = i$)
at zeroth order in the deformation.

The most general ansatz for the form of the metric after the
deformation, which preserves the $AdS_5$ isometries, is a warped
fibration of $AdS_5$ over a deformed ${S^5}'$, namely \be
\widetilde{ds}^2=  \exp(2 \rho) \, ds^2_{AdS_5} - ds^2_{S^{5'}}.
\ee Written in a 4-6 split of coordinates our ansatz for the 10d
metric is
 \bea \label{mansatz}
 \widetilde{ds}_{10}^{~2} &=& e^{2 \rho}\, \left[ r^2\,
\eta_{\mu\nu} \,dx^{\mu}
 dx^{\nu} - {1 \over r^2}\, ds^2 \right] = \non
 &=&  \fpar^2 \,\eta_{\mu\nu} \,dx^{\mu} dx^{\nu} - \fperp^2 \,ds^2,
\eea where $ \fpar \equiv r\, e^{\rho}, ~\fperp \equiv
e^{\rho}/r$,
 and we denote the 10d metric quantities with a tilde
and the 6d metric without one, $ds^2 = g_{\al \bt} dx^{\al}
dx^{\bt}$. At zeroth order $ds^2$ is the flat metric on $\IR^6$,
$ds^2_0=dr^2+r^2 \, ds^2_{S^5}$, and $\rho=0$.

The ansatz (\ref{mansatz}) has manifest 4d Lorentz invariance. In
order to guarantee the invariance under all the $AdS_5$
isometries we require that :
\bea ds^2(x^\al) &&~\mbox{is homogeneous of degree 2}, \non
 \rho(x^\al) &&~\mbox{is homogeneous of degree zero}, \non
g_{\al \bt}(x^\gamma)\, x^\bt &=&  x_\al \equiv \delta_{\al
\bt}\, x^\bt. \label{rfreeAnsatz} \eea
 The ansatz should also have a $U(1)$
isometry (corresponding to the $U(1)_R$ in the $\cn=1$
superconformal algebra) which we discuss later after introducing
complex coordinates.

Let us compute the connection $\tilde \omega$ for the metric
(\ref{mansatz}). The frame of 1-forms is \bea
 \tOmega^\hmu &=& \fpar \, dx^\mu, \non
 \tOmega^\hal &=& \fperp \, \Omega^\hal = \fperp \,
\Omega^\hal_{~ \bt} \, dx^\bt, \eea where hatted indices are in
orthonormal frames, and $\tOmega,\, \Omega$
 are the frames of 1-forms
  for ${\tilde g}, \,g$ respectively. A frame of 1-forms is defined by the
orthogonality relation $g=\Omega^\hal \, \Omega^\hbt \, \eta_{\hal
\hbt}$ where $\eta_{\hal \hbt}$ is the standard flat metric. We
will also use the frame of tangent vectors, $e_\hal$ (and
$\te_\hal$ for ${\tilde g}$), defined by $dx^\bt=e^\bt_\hgm
\,\Omega^\hgm$ namely $e^\bt_{~\hgm}$ is the inverse of
$\Omega^{~\hgm}_\bt$.

To find the connection we use Cartan's structural equations \bea
 d \Omega^\hM + \omega^\hM_{~\hN} \,\Omega^\hN =0, \non
 \omega_{\hM \hN} + \omega_{\hN \hM} = dg_{\hM \hN}=0.
 \eea

In the case of an $\IR^4$ index we have
 \be
d\tOmega^\hmu = d\fpar \, dx^\mu =\del_\al \fpar\, dx^\al
 dx^\mu =
{\del_\al \fpar \over \fperp}\, e^\al_\hbt\, \tOmega^\hbt \,
dx^\mu, \ee and the solution is \bea
 \tomega^\hmu_{~\hnu} &=& 0, \non
 \tomega^\hmu_{~\hbt} &=&
+{\del_\al \fpar \over \fperp}\, e^\al_{~\hbt} \,dx^\mu, \non
 \tomega^\hbt_{~\hmu} &=& - \eta^{\h{\bt}\h{\gm}}\,
 \eta_{\h{\mu} \h{\nu}}\, \omega^\h{\nu}_{~\h{\gm}}.
 \eea

The equations for an $\IR^6$ index are \bea
 d \tOmega^\hal &=& d\fperp ~\Omega^\hal + \fperp \,d\Omega^\hal = \non
 &=& -\partial_\bt \fperp \,\Omega^\hal \,dx^\bt -\fperp \,\omega^\hal_{~\hbt}
 \,\Omega^\hbt= -\fperp^{-1} \,\partial_\bt \fperp \,\Omega^\hal \,
e^\bt_{~\hgm}
 \,\tOmega^\hgm - \omega^\hal_{~\hbt} \,\tOmega^\hal, \eea
 and the solution is
 \be
\tomega^\hal_{~\hbt} = \fperp^{-1} \,\partial_\gm \fperp
(\Omega^\hal \,e^\gm_{~\hbt} - \Omega_\hbt \,e^{\gm \hal}) +
\omega^\hal_{~\hbt}. \ee

In order to evaluate the equations (\ref{susyvar}) we need the
expression for the spin connection. Using our results above, in
the $\IR^4$ directions :
 \bea
{1 \over 4} \,\tomega_{\mu \hM \hN} \,\Gamma^{\hM \hN} &=& {1
\over 2} \,\tomega_{\mu \h{\mu} \hbt} \,\Gamma^{\h{\mu} \hbt} =
\non
 &=&  {1 \over 2} \,{\del_\al \fpar \over \fperp}
\,e^\al_\hbt \,\Gamma_\h{\mu}^{~~\hbt} = \non
 &=& \half (r^2\, \del_\al \rho + x_\al ) \Gamma_\h{\mu}^{~\al}
 =\non
 &=& \half r\, \tGamma_{\h{\mu}}( r\, \del \Slash \rho +
 \Gammar )/\fpar
\label{omega_slash} \eea where the various gamma matrices are
defined as usual by $\{ \Gamma_M,\Gamma_N \}=2g_{MN}, ~\{
\Gamma_\hM,\Gamma_\hN \}=2\eta_{\hM \hN}, \{ {\tilde
\Gamma}_\al,{\tilde \Gamma}_\bt \}=2{\tilde g}_{\al \bt}$. In the
orthonormal frame $\Gamma=\tGamma$.
We define $\Gammar=x^\al \,\Gamma_\al / r$ , which satisfies
$(\Gammar)^2=+1$.

Equation (\ref{omega_slash}) suggests to define $\omega \Slash_4$
by \bea
 {1 \over 4} \,\tomega_{\mu \hM \hN} \,\Gamma^{\hM \hN} &=& \tGamma_\mu
\,\omega \Slash_4, \non
 \omega \Slash_4
 &=& \half\, e^{-\rho}(r \, \del \Slash \rho + \Gammar).
 \label{omega_slash4}
\eea

In the $\IR^6$ directions the spin connection is \be
 {1 \over 4}\,\tomega_{\al \hM \hN} \,\tl{\Gamma}^{\hM \hN} =
{1 \over 4}\,\omega_{\al \hM \hN} \,\Gamma^{\hM
 \hN} + {1 \over 4}\, (-\del_\bt \rho + x_\bt/r^2)(\Gamma^\bt\, \Gamma_\al
- \Gamma_\al\,
 \Gamma^\bt). \ee

\sbsection{Other fields}

We will take the 3-form field to preserve the $SO(4,2)$
invariance, namely only $G_{\alpha \beta \gamma}$ will be
non-zero, and it will be homogeneous of degree (-3) and obey
$x^{\alpha} G_{\alpha \beta \gamma} = 0$.

Given our ansatz for the metric, the form of the 5-form field
strength $F$ is uniquely determined by the equations $F=*F$,
$dF=-4i G \wedge G^* =0$ (the last equation follows from the fact
that we only turn on $G$ with indices in the five compact
directions), up to an overall constant $f_5$ :
 \be F=-f_5 e^{-5 \rho}(\tOmega^{\h{0}} \,\tOmega^{\h{1}} \,\tOmega^{\h{2}}
 \,\tOmega^{\h{3}} \,\tOmega^{\h{r}} +
 \,\tOmega^{\h{4}} \,\tOmega^{\h{5}} \,\tOmega^{\h{6}} \,\tOmega^{\h{7}}
\,\tOmega^{\h{8}}
  \,\tOmega^{\h{9}} \cdot \tOmega^{\h{r}}). \label{Fansatz} \ee
The $\cdot$ stands for contraction of the 1-form
$\Omega^{\h{r}}\equiv \Omega^{\h{\al}} \,x_\al/r$ with the 6-form
on its left through the metric, and the overall $(-)$ sign is
chosen so that the SUSY variation equations (through equation
(\ref{Fslash})) match those of \cite{GranaPolchinski} for
$f_5=1$. The constant $f_5$ is fixed by the normalization
condition, which in our conventions (including $R_0=\kappa=1$) is
given by $-\int_{S^5} F =1$. The following expressions will be
needed for the equations :
 \bea
 {i \over 480}  \,F \Slash &=& {f_5\over 4} e^{-5 \rho} \,\Gammar
\,(\Gamma^7 - \Gamma^4)
  \label{Fslash} \\
 1 = -\int_{\widetilde{S^5}} F &=& f_5 \int_{S^5}
 \sqrt{\mbox{det}(g_{\alpha \beta})}\, d^5 x
  \eea
  where we used equations (\ref{rfreeAnsatz}),(\ref{Fansatz}),
and $g_{\h{r}\h{r}}=1$.

We use the chirality matrix definitions\footnote{No confusion
should arise with the $\Gamma$ matrix associated with $x^7$ as we
will soon restrict ourselves to 6d.}
 $\Gamma^7=i \,\epsilon_{\al1 \dots \al6} \,\Gamma^{\al1}
\dots \Gamma^{\al6} /(6!)= i \,\Gamma^{4 \dots 9}$, $\Gamma^4=+i
\,\Gamma^{0123}$, which satisfy $\Gamma^7=- \Gamma^4 \,
\Gamma_{11}$ and thus conform with the sign conventions of
\cite{GranaPolchinski}. Since $\Gamma_{11} \,\eps= -\eps$ we have
$\Gamma^7 \, \eps= \Gamma^4 \, \eps$.

We take the ansatz for the Killing spinor to be \be
 \epsilon = \fperp^{-0.5} \chi =
 \fperp^{-0.5} (\zeta\, \chi^{(e)} + \zeta^*\, \chi^{(o)}), \label{eps_ansatz}
\ee where we explicitly factored out the dependence of $\epsilon$
on the $AdS_5$ coordinates. This form uses the 4-6 split and the
fact that the $\IR^4$ spinor part factors out of the equations,
except for complex conjugation, to decompose $\chi$ into the
product of an $\IR^4$ spinor and an $\IR^6$ spinor. $\zeta$ is an
arbitrary constant positive chirality $\IR^4$ spinor $\Gamma^4\,
\zeta = +\zeta, ~\Gamma^4\, \zeta^* =-\zeta^*$, and the
chiralities of the $\IR^6$ spinors are $\Gamma^7\, \chi^{(e)} = +
\chi^{(e)}, ~\Gamma^7\, \chi^{(o)} = - \chi^{(o)}$ where the
superscript $e/o$ stands for even/odd chirality (this will turn
out to agree also with the order in perturbation theory).

\subsection{Substitution of the ansatz into the equations}

Using the 4d Lorentz symmetry we can reduce all the equations to
6d, and since at zeroth order the untilded metric $ds^2$ is flat
$\IR^6$, we choose to work always in the untilded frame (with
positive $(+,\dots,+)$ metric). Due to 4d Lorentz symmetry the
equation for the gravitino components along $\IR^4$ can be
simplified into an algebraic equation for the spinor. Let us
collect the terms needed for this equation : \bea
 D_\mu -{i \over 2}\,Q_\mu &=& \del_\mu + {1 \over
4} \, \tomega_{\mu \hM \hN} \,\Gamma^{\hM \hN} =
 \tGamma_\mu \omega \Slash_4, \non
{i \over 480} \,F\Slash \,\tGamma_\mu  &=& - \tGamma_\mu\,
{f_5\over 4}\, e^{-5 \rho}\, \Gammar (\Gamma^7+\Gamma^4),  \non
-2 \, G\Slash \,\tGamma_\mu -\tGamma_\mu \,G\Slash &=& +
\tGamma_\mu \, G\Slash, \eea where we used (\ref{omega_slash4})
for $\omega_{\mu}$ and the fact that $\del_{\mu}=0$ for all
fields in the ansatz (including $Q_{\mu}=0$) in the first line,
(\ref{Fslash}) in the second, and $G \Slash \,\tGamma_\mu+
\tGamma_\mu \,G \Slash=0$ in the third. Assembling them, using
the ansatz for $\eps$ (\ref{eps_ansatz}), and the rescaled
variable $~ \delta \psi \equiv \fperp^{1.5} \,\tGamma^\mu
\,\delta \Psi_\mu /4$ we obtain
 \be
 \delta \psi = \left[ {1 \over 2 r}\, \Gammar\, (1- f_5\,
e^{-4 \rho}\, \Gamma^7)
  + \half\, \del \Slash \rho \right] \chi + {e^{\rho} \over 96\, r}
G \Slash \chi^*. \ee

Next, we collect the terms required for the gravitino equation
with an $\IR^6$ index : \bea \tl{D}_\al &=& D_\al + {1 \over 4}\,
(-\del_\bt \rho + x_\bt/r^2)
(\Gamma^\bt\, \Gamma_\al - \Gamma_\al\, \Gamma^\bt), \\
\fperp^{0.5} ( \del_\al \fperp^{-0.5} ) &=& \del_{\al}-{\del_\al
\fperp \over 2\, \fperp}= \del_{\al}+\half (-\del_\al \rho +
x_\al/r^2) \Rightarrow \non \Rightarrow \fperp^{0.5}\,
\tl{D}_\al\, \fperp^{-0.5} &=& D_\al +
{1 \over 2}\, (-\del \Slash \rho + \Gammar /r) \Gamma_\al, \\
{i \over 480} \,F\Slash \,\tGamma_{\al}  &=& -{f_5 \over 4\, r}\,
e^{-4 \rho}\, \Gammar\, \Gamma_\al\,  (\Gamma^7+\Gamma^4). \eea
 Assembling them, using the ansatz for $\eps$ (\ref{eps_ansatz}),
and the rescaled variable $~ \delta \psi_\al \equiv \fperp^{0.5}
\, \delta \Psi_\al $, we get \bea
  \delta \psi_\al &=& \left[ D_\al -{i \over
2}\,Q_\al +{1 \over 2\, r} \Gammar\, \Gamma_\al\, (1- f_5\,
e^{-4\, \rho} \Gamma^7) - \half\, \del \Slash \rho\, \Gamma_\al
\right] \chi \non && - {e^{\rho} \over 96\, r} (2 \, G\Slash
\,\Gamma_\al + \Gamma_\al \,G\Slash) \,\chi^*.
\label{vector-spinor} \eea

Once ``traced'', the vector-spinor equation (\ref{vector-spinor})
gives \be
 \Gamma^\al\, \delta \psi_\al = \left[ D \Slash -{i
\over 2}\,Q \Slash - (2/r)\, \Gammar\, (1-f_5 \, e^{-4 \rho}\,
\Gamma^7) + 2\, \del \Slash \rho \right] \chi - 6\, {e^{\rho}
\over 96\, r} G \Slash \chi^*, \ee where we used $\Gamma^\al\,
\Gamma^\bt\, \Gamma_\al= -4\, \Gamma^\bt$ and $\Gamma^\al\, G
\Slash\, \Gamma_\al = (-3 \cdot 2 + 6)\, G \Slash = 0$. This
equation can be combined with the spinor equation to create a
linear combination $\obst \equiv \Gamma^\al\, \delta \psi_\al +
6\, \delta \psi$ where all the $G \Slash \chi^*$ terms are absent.
This combination will play an important role later, and it is
given by \be \fbox{$ \obst= \Gamma^\al\, \delta \psi_\al + 6\,
\delta \psi = \left[ D \Slash -{i \over 2}\,Q \Slash + (1/r)\,
\Gammar\, (1-f_5 \, e^{-4 \rho}\, \Gamma^7)
 +5 \, \del \Slash \rho \right] \chi =0.~
$} \label{obst} \ee

After one additional rescaling $\lambda=-i\, \fperp^{0.5}\,
\Llambda$, we may list the full equations (\ref{susyvar}) in our
ansatz :
\be \begin{array}{|rcl|}
 \hline
 \room 0 = \delta \psi_\al &=& \left[ D_\al -{i \over 2}\,Q_\al +
 {1 \over 2\, r} \Gammar\, \Gamma_\al\, (1- f_5\, e^{-4 \rho}\, \Gamma^7)
- \half\, \del \Slash \rho\, \Gamma_\al \right]
  \chi ~ \\
 \room & & - {e^{\rho} \over 96\, r} (2 \, G\Slash \,\Gamma_\al +
\Gamma_\al \,G\Slash) \,\chi^*, \\
 \room  0 = \delta \psi &=& \left[ {1 \over 2 r}\, \Gammar\,
(1- f_5\, e^{-4 \rho}\, \Gamma^7) +
 \half\, \del \Slash \rho \right] \chi + {e^{\rho} \over 96\, r}
G \Slash \chi^*, \\
 \room 0=\delta \lambda &=& P\Slash \,\chi^* -{4 \over 96} \,G\Slash \,\chi. \\
 \hline \end{array} \label{eqset}
\ee When we use the 4-6 split of the spinor, each equation above
actually decomposes in two according to the $\zeta$ and $\zeta^*$
components in the ansatz (\ref{eps_ansatz}).

The $Q_\al$ term will not be needed for our calculations since it
will turn out to be quartic in the perturbation parameters $h$ :
$Q \sim B\, B^*, ~ B \sim h^2$, and we will only work to third
order in $h$.

At zeroth order the equations reduce to\footnote{These equations
are the same as in \cite{GranaPolchinski} since  $\Gamma^4 \chi
=\Gamma^7 \chi$.}
 \bea
 (\delta \psi_\al)^{(0)} &=& \del_\al \,\chi^{(0)} +  \Gammar \,\Gamma_\al \,
{1-\Gamma^7 \over 2r} \,\chi^{(0)} = 0,\non
 (\delta \psi)_0 &=& \Gammar \,{1-\Gamma^7 \over 2r} \,\chi^{(0)} = 0,\eea
from which we see that all 4 constant, positive chirality $\IR^6$
spinors $\Gamma^7 \chi^{(0)} = + \chi^{(0)}$ are solutions. Since
we can combine each of these spinors with 2 independent positive
chirality $\IR^4$ spinors, we find 8 complex spinors, or 16 real
supercharges (we discuss here only the supercharges which
correspond to four dimensional supersymmetries, as opposed to
superconformal transformations).

\subsection{The linearized equations, zero mode and obstruction}
\label{sec_linear}

Our method to find nearby solutions is to expand the fields and
Killing spinor as a series in a perturbation parameter $h$ and to
solve them order by order. To begin, we need to discuss the small
(linearized) fluctuations around the $AdS_5\times S^5$ solution.
The spectrum of these was found by Kim, Romans and
van-Nieuwenhuizen \cite{Kim:1985ez}. Since we want to preserve
the isometries of $AdS_5$, our solution will only involve turning
on AdS scalars, whose spectrum is summarized in figure
\ref{KRvNfig}.

Since we require that only the minimal amount of supersymmetry be
preserved (4d $\cn=1$ superconformal) we choose one specific
spinor $\chi^{(0)}=\chi_0$ out of the 4 (positive chirality)
$\chi$'s, to be the zeroth order spinor corresponding to the
preserved supercharges. This choice breaks
 the global symmetry
$SO(6)_R \to U(1)_R \times SU(3)$, with ${\bf 4} \to {\bf 3}_1 +
{\bf 1}_{-3}$.
 It is convenient to use complex coordinates in
$\IR^6 \simeq \IC^3$, replacing the $SO(6)$ indices
$\al,\bt,...$, by $SU(3)$ indices $i,j,\bi,\bj$. We choose the
flat metric to be of the form $ds_0^2= 2 \,dz^i \,d\bz^\bi$,
namely we normalize the complex coordinates such that $\eta_{i
\bi}=\eta_{ \bi i}=1$ and other components vanish. This
normalization allows us to do away with all barred indices --
they can always be lowered or raised by $\eta$ and become
unbarred with no additional factors (the price being the factor
of 2 in the expression for $ds^2_0$).

Having chosen a complex structure, we can identify the gamma
matrices with creation and annihilation operators. We will use
lower case gamma matrices $\gammad_j,\gammad^i$ to denote the
gamma matrices in the complex coordinates, satisfying $ \{
\gamma^i,\gammad_j \} = 2 \, \delta^i_j$, and we will choose our
coordinates such that $\chi_0$ is identified with the Clifford
vacuum, $\gammad_i \, \chi_0=0$ (and by complex conjugation,
$\gamma^j \chi_0^* = 0$). The dagger on $\gamma^\dag_j$ with
lower indices is a redundant notation to remind us that
$\gammad_j$ is a creation operator with respect to the ``state''
$\chi^*_0$. Using the complex coordinates we can easily formulate
the requirement for $U(1)_R$ isometry by requiring that each term
in the expansion of the fields has an equal number of $z$'s and
$\bz$'s\footnote{Through the Hopf fibration we can think of $S^5$
as a circle fibration over ${\bf CP}^2$. The fact that the
$U(1)_R$ isometry stays intact throughout the deformation means
that the problem can be described as a deformation of ${\bf
CP}^2$ rather than $S^5$, but we will not use this description
here.}.

Next, we will linearize our equations (\ref{eqset}) around the
$AdS_5\times S^5$ solution. This will give us the equations for
the leading deformation around the original solution, and at
every order $n$ in the perturbation series we will get the same
form of equations for the $n$'th order fields, with additional
contributions coming from the lower order fields. At the
linearized level the metric is flat, and so $\Gamma_\al \simeq
\Gamma_\hal$. The equations (\ref{eqset}), and therefore also the
linearized equations,
 may be split into components whose 4d
spinor is proportional to $\zeta$ or to $\zeta^*$.
 This decomposition turns out to distinguish odd and even orders since
the perturbation parameters $h$ (which correspond to a particular
mode of the 3-form field $G$) are odd under the element $(-{\bf
1}) \in \SLt{\IZ}$ which inverts $G$ and keeps the other bosonic
fields invariant, and the original $AdS_5\times S^5$ solution is
invariant under this $\IZ_2$. The $\IZ_2$ is actually a $\IZ_4$
when it acts on the fermionic fields, multiplying them (and the
spinor $\chi$) by $\pm i$ (depending on their $\Gamma_{11}$
chirality); this has the same effect as taking $\chi \to \chi$,
$\chi^* \to -\chi^*$. This implies that at odd orders the only
bosonic field that can be turned on is $G$, and the only
fermionic modes that will be turned on are $\chi^{(o)}$ obeying
$\Gamma^7\, \chi^{(o)}=- \chi{(o)}$, while at even orders
$g$,$B$, and $\rho$ get contributions, as well as $\chi^{(e)}$
which obeys $\Gamma^7\, \chi^{(e)}=+ \chi^{(e)}$. The linearized
equations at odd orders are thus of the form : \bea
 0=\delta \psi_\al &=& (\del_\al+ \Gammar \, \Gamma_\al/r)
 \chi^{(o)}
 -{1 \over 96\, r} (2 \, G\Slash \, \Gamma_\al +\Gamma_\al \, G\Slash)
\chi^*_0 + \dots, ~ \label{odd1}
 \\
\room 0= \delta \psi &=& {1 \over r}\, (\Gammar \, \chi^{(o)} +
{1 \over
96} \, G\Slash \, \chi^*_0) + \dots, ~ \label{odd2} \\
 \room 0= \delta \lambda &=& -{4 \over 96} \, G\Slash \, \chi_0 +
\dots, ~ \label{odd3} \eea where the $\dots$ refers to
contributions involving fields of lower order in the deformation
parameter.

Similarly, the linearized equations at even orders are : \bea
 0=\delta \psi_\al &=& \del_\al \chi^{(e)} + {1 \over 4} \,
\omega\Slash_\al \, \chi_0 + \half \,  [-\del \Slash \rho + (4
\rho - \delta f_5)/r ~\Gammar]\,
\Gamma_\al \chi_0 + \dots, \label{even1} \\
 0 = \delta \psi &=& {1 \over 2} \,[r \del_\al \rho \, \Gamma^{\hal}
 + (4 \rho -\delta f_5) \,\Gammar] \,\chi_0 + \dots, \label{even2}
 \\
 0 = \delta \lambda &=& \del \Slash B \, \chi_0^* + \dots, \label{even3} \eea
where we defined $f_5 = 1 + \delta f_5$, and we also need to
impose the flux quantization constraint
 \be 0=\delta f_5 + \int_{S^5}  \mbox{tr}(g_{\alpha \beta}
-\eta_{\alpha \beta}) \, d^5 x + \dots.   \ee

For later use we linearize $\obst$ as well. At odd orders \be
 \obst =( \del \Slash + {2 \over r}\, \Gammar ) \chi + \dots \ee
 While at even orders \be
 \obst = \del \Slash\, \chi+ \left[ \omega \Slash_6 -{i \over 2}\,Q \Slash
 + (1/r)\, \Gammar\, (4\, \rho -f_5 )
 + 5\, \del \Slash \rho \right] \chi_0 \dots \ee
where $\omega \Slash_6=\omega \Slash_\al\,  \Gamma^\al$.

The odd equations have a zero mode that will turn out to be
crucial for our analysis, and we will describe it in detail in
subsection \ref{obstthird}.

\subsection{First order}

Now, we have everything we need in order to perform an order by
order analysis of the equations. A field $\phi$ at order $(n)$
will be denoted by $\phi^{(n)}$.

The first order analysis was performed already in
\cite{GranaPolchinski}. From the spectrum of scalars we know that
the only massless ones are the ${\bf 1}_\IC$ from $B$, the ${\bf
45}_\IC$ from $G$ and the ${\bf 105}_\IR$ from the metric and
5-form.
Out of these marginal modes the only one (except for $B$) which
is super-marginal (namely, supersymmetric and marginal) is in the
${\bf 10}_0$ representation, which appears in the ${\bf 45}$ when
we decompose it under $SO(6) \to SU(3) \times U(1)_R$ (see
appendix B for our group theory conventions). This mode comes
from $G$ so it is odd under the $\IZ_2$ discussed above. To
parameterize the coefficient of this mode we introduce the
perturbation parameters $h_{ijk}$, which are a 3rd rank symmetric
tensor of $SU(3)$. $h$ will be proportional to the superpotential
deformation in the field theory dual, and we will determine the
precise relation between them in the next section.

At this point we specialize our notation to the perturbation
problem. As mentioned already we work in the untilded frame where
the metric is flat at zeroth order $\eta_{i \bj}= \delta_{ij}$,
so $r^2=2\, z^i \bz_i$. All indices are raised and lowered by
means of $\eta_{i \bj}$, and all gamma matrices are expressed by
lower case $\gamma$'s
which are defined in an orthonormal frame. We define $h$ tensors
with less than 3 indices by contracting with $z's$, namely
$h_{ij} \equiv h_{ijk} \,z^k, ~h_{i} \equiv h_{ijk}\, z^j\, z^k,
~h\equiv h_{ijk}\, z^i\, z^j\, z^k\, , \bh^{ij} \equiv \bh^{ijk}\,
\bz_k\, , \dots$ (where $\bh^{ijk}$ is the complex conjugate of
$h_{ijk}$). Indices are antisymmetrized without any prefactor,
$h_{[i} \bz_{l]}=h_i\, \bz_l-h_l\, \bz_i$. We record here a
useful relation $\gammad_{123} \, \gamma^{123} \, \chi_0 = -8
~\chi_0$. We also define \be \gammar=(\bz_i\, \gamma^i + z^i\,
\gammad_i)/r. \ee

The explicit form of $G^{(1)}$ in terms of $h$, corresponding to
turning on the massless scalar in the $\bf{10}_0$ representation,
is quite involved. It can be deduced from the expressions for a
general superpotential in \cite{GranaPolchinski} to be
 \bea {1 \over 96}\, G^{(1)} &=& {1\over 6}
h\, \epsilon^{ijk}\, ~d\bz_{ijk} /r^6  \non
 &-& ({1\over 2}r^2\, h_{il} + h_{[i}\, \bz_{l]})\, {1\over 2} \epsilon^{jkl}
\, ~d\bz_{jk}\, dz^i
 /r^6 \non
 &+& h_{il}\, \bz_j\, \bz_m\, \epsilon^{lmk}\, dz^{ij}\, d\bz_k /r^6,
 \label{G1} \eea
where a factor of $1/96$ was inserted in the normalization of $h$
for later convenience. This expression can be derived by noting
that from equation (\ref{odd3}) we get \be 0=G\Slash\, \chi_0=
6\, G_{123}\, \gamma^{123}\, \chi_0 + G^i_{~jk}\, (2\, \gamma_i\,
\gamma^{jk} +\gamma^{[k}\, \gamma_i\, \gamma^{j]} +2\,
\gamma^{jk}\, \gamma_i)\, \chi_0, \ee and since both terms have to
vanish we get $G_{123}=0, ~G^j_{~ij}=0$. Similarly one continues
to analyze the other equations for $G$ to arrive at (\ref{G1}).

Slashing  $G^{(1)}$ one gets \be \begin{array}{|rcl|}
 \hline
 ~\rule[-3mm]{0mm}{9mm} ~~{1 \over 96} r^3\, G\Slash^{(1)} &=& {1\over 6}\, h\,
\epsilon^{ijk}\, \gammad_{ijk} \\
 ~\rule[-3mm]{0mm}{9mm} &-& ({1\over 2} r^2\, h_{il} + h_{[i}\, \bz_{l]}) {1\over 2}\epsilon^{jkl}\,
 \gammad_{jk}\, \gamma^i  + 2\, h_i\, \bz_j\, \gammad_k\,
 \epsilon^{ijk} ~~
 \\
 ~\rule[-3mm]{0mm}{9mm}&+& h_{il}\, \bz_j\, \bz_m\, \epsilon^{klm}\, \gammad_k\,
 \gamma^{ij}. \\
\hline \end{array} \ee
 The expression for $\chi^{(1)}$ can be
found from equation (\ref{odd2}),
 \be \fbox{$\displaystyle ~\chi^{(1)}=- {1 \over 96}\, \gammar\, G\Slash^{(1)}\, \chi_0^* =
-{1 \over 2}\, h_i\,
 \gamma^i\, \gammad_{123}\, \chi_0^* /r^2. ~$} \label{chi1} \ee
  Note that $\Gamma^7\, \chi^{(1)} = - \chi^{(1)}$ and that equation
(\ref{odd1}) is also automatically obeyed.

\subsection{Second Order}

We now continue to expand the equations into second order. Here
we found it convenient to perform our computations both by hand
and by computerized symbolic computations using {\it Mathematica}
\cite{mathematica}. The computer computation required
implementing the Clifford algebra $\gamma_i$ as well as the other
tensors which appear here $h_{ijk}, \epsilon_{ijk}, \delta^i_j$
(see \cite{GAMMA} for a related {\it Mathematica} package).

We start with $B^{(2)}$. From the third equation in (\ref{eqset})
(whose linearized form is (\ref{even3})) we have at second order
$\del \Slash B^{(2)}\, \chi_0^*={4 \over 96} (G \Slash \,
\chi)^{(2)}
= {4 \over 96} G \Slash^{(1)} \, \chi^{(1)}
$. We compute \be
 {1 \over 96} G \Slash^{(1)} \, \chi^{(1)}=-{1 \over 96^2}
\Gsla \, \gammar \, \Gsla \, \chi_0^* = -4 \, h_{ij}  h_k \, \bz_m
 \bz_n \, \epsilon^{ikm} \, \epsilon^{jln} ~ \gammad_l \, \chi_0^*
/r^5. \ee
 The resulting differential equation is integrable and we get
 \be \fbox{$\displaystyle  B^{(2)}=-2 ~h_{ij} h_{kl} ~\bz_m
\bz_n ~\epsilon^{ikm} \epsilon^{jln}/r^4.$ } \ee
This expression is in the ${\bf 27}$ representation of $SU(3)$,
which includes the modes with no $U(1)_R$ charge coming from the
spherical harmonics of $B$ in the ${\bf 105}$ representation of
$SO(6)$. This can be seen from the fact that as a quadratic
symmetric form in $h$ it could apriori be in the $\left( {\bf 10}
\times {\bf 10} \right)_s={\bf 28} + {\bf 27}=[6,0]+[2,2]$
representation, but the ${\bf 28}=[6,0]$ is the symmetrization on
all 6 indices which vanishes for our expression (due to the
$\eps$'s), so it must be proportional to the ${\bf 27}$.

Turning next to $\rho$, we note that it appears in the linearized
equations only in the combination $\rho' \equiv \rho-\delta
f_5/4$. Moreover, equations (\ref{even2}),(\ref{eqset}) give us
an equation only for $\rho'$ (recall that $f_5$ is a constant) :
\be {1\over 2}(4 \rho'\, \gammar\, + r\, \del \Slash
\rho')=-{1\over 96} \, G\Slash^{(1)}\, \chi^{(1)*} =
 {1\over {96^2}} \, G\Slash^{(1)}\, \gammar\, G\Slash^{(1)*}\, \chi_0^*.
\ee
Using our first order results this is given by  \be {1 \over
96^2} \bGsla \, \gammar \, \Gsla\, \chi_0^* = 2 \, h_i \bh^{il}
\, \gammad_l \, \chi_0^*. \ee
 The resulting first order partial differential equation for $\rho'^{(2)}$,
\be 4 \, \rho'^{(2)} \, \bz_m /r^2 + \del_m \rho'^{(2)}=4 \,
h_{mj} \bh^j /r^4, \ee
  is integrable and yields
 \be  \fbox{$\displaystyle
\rho'^{(2)}=\rho^{(2)}-{1\over 4}\delta f_5^{(2)}=  2 ~h_i ~\bh^i
 /r^4. $} \ee
Note that this expression is proportional to $V=|\del W|^2$ for
$W \propto h_{i j k} \, z^i z^j z^k$, which is the second order
contribution to the field theory potential due to our deformation
of the superpotential, and that the mode of $\rho'$ in the $\bf
105$ spherical harmonic is precisely the massless field which is
identified with the fourth order scalar potential deformation in
the field theory. Thus, supergravity automatically turns on the
second order deformations in the field theory which are implied
by the first order deformation and by supersymmetry.

The equation for the perturbations in the metric and the Killing
spinor can now be deduced from equation (\ref{even1}),
 \bea \label{metrictwo}
\del_\eps \chi^{(2)} + \del_{[\al} S_{\bt]\, \eps}^{(2)} \,
\gamma^{\al \bt}\, \chi_0 &=& B_\eps =
 \non
 &=& {2 \over 96} \Gsla  \, \gamma_\eps \, \chi^{(1)*} -
(\del_\eps \rho'^{(2)} + 4\,
 \rho'^{(2)}\,  z_\eps/r^2) \chi_0 + \del \Slash \rho\, \gm_\eps\, \chi_0,
\eea
where we defined $S$, a 2nd rank symmetric and real tensor, to be
the correction to the frame and metric, \bea
\Omega^{\hat{\alpha}}_{~\beta}=\delta^{\hat{\alpha}}_{~\beta}-4\,
S_{\alpha \beta}, \label{defS} \non
 \fbox{ $g_{\al \bt}=\eta_{\al \bt} -8 S_{\al \bt}$, } \eea
 and we defined
 \bea
B_m & = &-16\, h_{[m}\, \bz_{p]}\, \bh^p\, \chi_0 /r^6 +8\,
h_{i[m} \, \bz_{p]}\, \bz_j\, \bh^p\, \gamma^{ij}\, \chi_0 /r^6
-4\, h_{j m}\, \bh^j\,
 \chi_0/r^4,\non
 B^m & = &16\, \bh^m\, h\, \chi_0/r^6
 - 8\, h_i\, \bz_j\, \bh^m\, \gamma^{ij}\, \chi_0/r^6
  + 2\, \del_j( h_i\, \bh^i/r^4) \gamma^{jm}\, \chi_0
  + 4\, h_j\, \bh^{jm}\,  \chi_0/r^4 \non
&& -16\, h_i\, \bh^i\, z^m\, \chi_0/r^6.
 \eea

We find that the $\gamma^{ij} \, \chi_0$ terms in
(\ref{metrictwo}) can be matched by
 \be \begin{array}{|rcl|}
 \hline
 \room S_{ij}^{(2)} &=& h_{ij}\, \bh /r^4 - 2\, \bz_i \bz_j\,
h_p \bh^p /r^6 +
\del_{i} \del_{j} \hphi,~\\
 \room S^{ij(2)} &=& {\overline {S_{ij}^{(2)}}}, ~\\
 \room S^{i~(2)}_{~j} &=& -\bh^i\,  h_j /r^4 + \delta^i_{~j}\, h_k\,
\bh^k/r^4 + {1\over 8}(h_{jkl}\, \bh^{ikl} - {1\over 3}
\delta^i_j\, h_{klm} \bh^{klm}),  \\
 \hline \end{array}
\ee where $\hphi$ is to be determined, and without any
contribution involving $\chi^{(2)}$. Our supersymmetry equations
depend only on derivatives of $S$, so we can always add arbitrary
constant terms to $S$, as we did in our expression for $S^i_j$;
the reason for adding the particular constant terms that we added
will be explained below. We record for later use that \be
S^{j~(2)}_j=2\, \bh^j\, h_j/r^4. \ee

The $\chi_0$ term in (\ref{metrictwo}) now implies \bea
 \del_m (\h{\chi}^{(2)} + 2\, \del^j \del_j \hphi) =
  \del_m(-5\, h_i\, \bh^i/ r^4), \non
  \del^m (\h{\chi}^{(2)} - 2 \del^j \del_j \hphi) =
\del^m(5\, h_i\, \bh^i/ r^4), \eea where we defined $\chi^{(2)}=
\h{\chi}^{(2)}\, \chi_0$.
 This leads to
 \be \label{fork}
2\, \del^j \del_j \hphi = \Delta \hphi = -5\, h_j \bh^j/r^4, \ee
 \be
\fbox{ $\chi^{(2)} = 0 $. }
\ee
The fact that $\chi^{(2)}$ vanishes is somewhat surprising, and
could be a hint about a general feature of the full solution.

Before solving explicitly (\ref{fork}) for $\hphi$ let us
describe in some detail the general solution to a Poisson
equation, $~\Delta \hphi = \psi$, where $\psi$ is some source
function independent of the radial coordinate $r$. Decomposing
$\psi$ and $\hphi$ into spherical harmonics\footnote{The index
$m$ runs over all the states within the representation $l$.}
$Y_{lm}$ on $S^5$, by writing $\hphi = \sum_{lm} \hphi_{lm}(r)
Y_{lm}$, $\psi = \sum_{lm} \psi_{lm} Y_{lm}$, we find \be \sum
Y_{lm}\, [\del_r^2 + {5 \over r} \del_r - {1 \over r^2}\,
l\,(l+4)]\, \hphi_{lm} = \Delta \hphi = \psi= \sum \psi_{lm}\,
Y_{lm}. \ee
 The solution is \be \hphi_{lm} = \left\{
\begin{array}{l@{\quad:\quad}l}
 \psi_{lm} ~ r^2  /(12- l(l+4)) & l \neq 2, \\
 \psi_{lm} ~ r^2 \log(r/r_0)/8       & l=2.
 \end{array} \right. \ee
Note that the $l=2$ mode
stands out as it generates logarithms in $\hphi$. The explicit
solution of equation (\ref{fork}) for $\hphi$ is
 \be \fbox{$\displaystyle \hphi={1 \over 4}\, h_i \bh^i /r^2-{1 \over 48}
 (6\, h_{ij} \bh^{ij}-h_{ijk} \bh^{ijk}\, r^2)\log(r^2/r_0^2)-{1 \over
 6}\, h_{ij} \bh^{ij},$} \ee
for arbitrary $r_0$ (the $log(r_0)$ term multiplies a zero mode
of the Laplacian).

Let us check whether this is consistent with the conformal
isometries. The first condition in (\ref{rfreeAnsatz}) is
satisfied, as it is easy to check that the logarithm does not
appear in $\del_i \del_j \hphi$ which appears in the metric. We
require also that the metric satisfies the second constraint of
(\ref{rfreeAnsatz}). The linearized constraint is \be
 S_{i\al}\, x^\al = 0 \ee
 (a similar equation with upper index $i$ is the complex conjugate of
this one).
 We find
\bea
 S_{i\al}\, x^\al &=& S_{ij}\, z^j + S_i^{~j}\, \bz_j = \non
 &=& h_i \bh/r^4 - \bz_i\, h_p \bh^p /r^4 +  z^j\, \del_{i} \del_{j}\hphi -
 \non
 && -h_i \bh/r^4 + \bz_i\, h_p \bh^p /r^4 +
{1\over 8}(h_{ikl} \bh^{jkl} -{1\over 3}\delta_i^j h_{klm}
\bh^{klm}) \bz_j \non &=&  z^j\, \del_{i} \del_{j}\hphi +{1\over
8}(h_{ikl} \bh^{kl} -{1\over 3}\bz_i h_{klm} \bh^{klm}), \non
  z^j\, \del_{i} \del_{j}\hphi  &=& -z^j\, \del_{i} \del_{j} [{1\over 48}\,
(6\, h_{kl} \bh^{kl}-h_{klm} \bh^{klm}\, r^2)
 \log(r^2/r_0^2)] = \non
 &=& -{1\over 48}\, (6\, h_{ikl} \bh^{kl}-2\, h_{klm} \bh^{klm}\, \bz_i).
 \eea
We see that exactly for the choice we made of the constant terms
in $S^i_j$, this equation vanishes identically and our solution
is conformally invariant and supersymmetric. This agrees with our
field theory discussion in section \ref{FT}, where we argued that
if one swallows the wave function renormalization into the
normalization of the chiral superfields, we expect to find a
conformal solution to second order in the deformation whether or
not $(h \bh)_{\bf 8}$ vanishes.

It is interesting to note that the constant terms we added to
$S^i_j$ are exactly proportional to $(h \bh)_{\bf 8}$. These
terms seem to be related to a diffeomorphism of the form $z^i \to
z^i + {1\over {16}} [(h \bh)_{\bf 8}]^i_j z^j$; if instead we
would perform the same diffeomorphism with an imaginary
coefficient, this would correspond (from the $AdS_5$ point of
view) to an $SU(3)$ gauge transformation
with a gauge parameter $(h \bh)_{\bf 8} \log(r/r_0)$. This
suggests that these terms are related to a complexified $SU(3)$
transformation, and we saw at the end of section \ref{FT} that in
the field theory these transformations are precisely related to
the wave function renormalization induced by the anomalous
dimensions. Thus, at this order we seem to find a nice agreement
between the field theory and supergravity results, assuming that
supergravity corresponds to the field theory where we rescaled
the fields to have canonical kinetic terms. We will make this
more precise at third order, when we will compute the running of
the coupling constant (which is a physically measurable quantity
that can be directly compared between field theory and
supergravity).

We can also solve for the perturbation in the normalization
constant $\delta f_5^{(2)}$, although it will not be necessary
for any other computation presented in this paper. The flux
normalization equation at second order is
 \bea 0 &=& \delta f_5^{(2)} - \int_{S^5}  8\, S_j^{j(2)}\, d^5x \Rightarrow
\non
 \Rightarrow \delta f_5^{(2)} &=& 16 \int_{S^5}  h_i\, \bh^i/r^4\,
 d^5x = {2 \over 3} |h|^2, \eea
 where we used the definition of $S$ (\ref{defS}),
$\mbox{ tr}(S)=2\, S^j_{~j}$ and the integration formula
(\ref{s5integ}), and we defined $|h|^2\equiv h_{ijk}\, \bh^{ijk}$
which is a $z$ independent constant.

\subsection{$\obst$ at third order}
\label{obstthird}

In equation (\ref{obst}) we introduced the linear combination of
SUSY variation equations
\be
\obst= \Gamma^\al\, \delta \psi_\al + 6\, \delta \psi =
\left[ D \Slash -{i \over 2}\,Q \Slash + (1/r)\, \Gammar\, (1-f_5
\, \exp(-4\, \rho)\, \Gamma^7)
 +5\, \del \Slash \rho \right] \chi =0,~
\ee
where $D \Slash = \del \Slash + \tomega \Slash_6,~ \tomega
\Slash_6={1 \over 4}\, \omega_{\alpha,\hat{\beta}\hat{\gamma}}\,
\Gamma^\alpha\, \gamma^{\hat{\beta}\hat{\gamma}}$.

We will be interested in solving this equation at third order in
the deformation, where this term will be responsible for
eliminating the marginal but not exactly-marginal deformations,
but let us consider first the lower orders, including the even
orders. At zeroth order we have \be 0=\obst^{(0)}= \del \Slash
\chi_0, \ee which is clearly satisfied. At first order we find
 \be 0=\obst^{(1)}= (\del \Slash + {2 \over r} \gamma_r)
 \chi^{(1)}. \label{obst1} \ee
 This is an $SO(6)$ invariant equation which requires $\chi^{(1)}$
to be a zero mode of the differential operator $\diff$. This
operator has a single zero mode which is in the in the
$\overline{\bf 60}=[0,2,1]$ representation of
$SO(6)$.
The $\overline{\bf 60}$ includes two $SU(3)$ irreducible
representations with the correct R-charge (+3), which are in the
$\bf 8$ and $\bf 10$ representations of $SU(3)$. We can write
them down explicitly and check that they solve (\ref{obst1}) :
\bea
 (ZM{\bf 8})^i_j &=& (12\, z^i \bz_j/r^2 -2\, \delta^i_j)\, \chi_0^* -
 \epsilon_{mnj}\, z^i z^n\, \gamma^m \gammad_{123}\, \chi_0^*/r^2, \non
 (ZM{\bf 10})^{ijk} &=&
z^{(i} z^j\, \gamma^{k)} \gammad_{123}\, \chi_0^*/r^2. \eea In
our solution above, $\chi^{(1)}$ (\ref{chi1}) was precisely
proportional to $h_{ijk} (ZM{\bf 10})^{ijk}$, as required by
(\ref{obst1}). Let us also record these quantities after being
multiplied by $\gamma_r$ :
 \bea \label{badmodes}
\gamma_r (ZM{\bf 8})^i_j &=& (16\, z^i \bz_j z^k- 2\, z^i
\delta^k_j-2\,z^k \delta^i_j)
 \gammad_k\, \chi_0^*, \non
 \gamma_r (ZM{\bf 10})^{ijk} &=& {2 \over r^3}\, (z^{(i} z^j z^{k)}\,
 \gammad_{123}\,  \chi_0^* + 2\, \epsilon^{pq(i}\, z^j z^{k)}
\bz_p \,\gammad_q \, \chi_0^*). \eea

At second order we find
 \be 0=\obst^{(2)}= \del \Slash \chi^{(2)} + \left[ \tomega
 \Slash_6 + 5\, \del \Slash \rho + 4\, \rho'\, \gamma_r/r
 \right]^{(2)} \,\chi_0. \ee
This expression is consistent with our previous results $
~\tomega\Slash_6 =-4 \rho' \, \gamma_r/ r-5\, \del \Slash
\rho^{(2)}$, which follows from the expression for $S^{(2)}$, and
$\chi^{(2)}=0$ .

At third order we have
 \be \label{obst3}
0=\obst^{(3)}= (\del \Slash + {2 \over r} \gamma_r) \, \chi^{(3)}
+ \left[ \tomega
 \Slash_6 + 5\, \del \Slash \rho - 4 \rho' \,
\gamma_r/r \right]^{(2)} \chi^{(1)} + St, \ee where
 the term $St$ is defined by
\be St = 4\, S^{\alpha \beta}\, \gamma_\beta\, (\del_\alpha + 2
 \del_\alpha r /r )\, \chi^{(1)}.
\ee An explicit calculation gives
 \bea
St = (16\, h_i\, h_j\, \bh^{ij} - 12\, h_{ij}\, \bh^{ij}\, h -{2
\over 3}\, h\, h_{ijk}\, \bh^{ijk})\, \gammad_{123}\,
\chi_0^*/r^6 + \non
 (- 16\, h_i\, h_{jk}\, \bh^k\, + 2 r^4 \bh^{kmn} h_{imn} h_{jk})
\epsilon^{i j l}\, \gammad_l\, \chi_0^*/r^6.
 \eea

As we saw above, the differential operator $\diff$ has a zero
mode and its image is not the full space. We will denote \be
\label{obstdef} \obst^{(3)} = (\del \Slash + {2 \over r}
\gamma_r) \, \chi^{(3)} + Obst, \ee and if $Obst$ will have a
component which is not in the image of the differential operator
$\diff$ (acting on spinor spherical harmonics on $S^5$), the
equation will have no AdS-invariant solution. Since we can choose
a basis of spherical harmonics which are eigenfunctions of
$\diff$ up to multiplication by $\gamma_r$ (namely, they obey
$\diff \chi = \alpha \gamma_r \chi / r$), the requirement is that
$Obst$ will have no components proportional to the spherical
harmonics appearing in (\ref{badmodes}).

Substituting the expressions for $St$ and $\rho^{\prime(2)}$ into
(\ref{obst3}) we can evaluate $Obst$ and $\gammar Obst$ which
will be useful shortly : \bea
 r \, Obst &=&(16\, h_i h_j \bh^{ij}/r^5 + 16\, h h_k \bh^k/r^7
 -12\, h h_{ij} \bh^{ij} /r^5 -{2 \over 3}\,h |h|^2/r^3)\,
\gammad_{123}
 \chi_0^*+  \non &&
+ (16\, h_{ik} \bh^{k} h_j /r^5- 32\, \bz_i\, h_j h_k \bh^k/r^7 +
2 \bh^{kmn} h_{imn} h_{jk} / r)
\epsilon^{ijl} \,\gammad_l \chi_0^*,  \\
 r\, \gammar\, Obst &=& (- 32\, h_i h_{jk} \bh^k + 4 r^4 \bh^{kmn}
h_{imn} h_{jk}) \bz_l\, \epsilon^{ijl} \chi_0^*/r^6 + \non
 &&\bigg[16 h_j h_k \bh^{jk} \bz_i + 8 h\, h_{ij} \bh^{j} - 12 h h_{jk}
 \bh^{jk}\, \bz_i -{2 \over 3}\, r^2\, h\, |h|^2 \bz_i + \non
&& r^4 \bh^{klm} h_{ilm} h_k - r^4 \bh^{klm} h_{lm} h_{ik} \bigg]
\gamma^i
 \gammad_{123}\, \chi_0^*  /r^6.
 \eea

\subsection{Projections of $Obst$ and the obstruction to superconformality}

In order to check whether we can solve (\ref{obst3}) with a
conformally invariant solution we need to project the spinor
$Obst$ onto the modes (\ref{badmodes}), and check whether this
projection vanishes or not. The equations for the vanishing of
this projection give two cubic equations for $h$ (actually $h\,
h\, \bh$), one in the ${\bf 10}$ representation and the other in
the ${\bf 8}$. In order to match with the field theory both of
these two equations must have $(h \bh)_{\bf 8}$ as a factor. For
the cubic ${\bf 8}$ this turns out to be trivial since there is a
unique cubic expression in this representation, which factorizes
(as shown explicitly below). For the ${\bf 10}$ the situation is
more interesting: two different expressions can be formed out of
$h$ and only a single linear combination factorizes. As we will
see in this section, it is exactly this combination that happens
to appear in the projection of $Obst$. This is a non-trivial test
of the AdS/CFT correspondence.

In general the projection of a spinor $\chi$ onto a mode $\chi_i$
is given by the inner product
\be (\chi_i,\chi)=\int_{S^5} \bar{\chi_i}\,  \chi. \ee
 A useful formula is
 \be \int_{S^5} z^{i_1} \dots z^{i_n}\,  \bz_{j_1} \dots
\bz_{j_n} = \sum_{\sigma \in S_n} \delta^{i_{\sigma(1)}}_{j_1}
\dots \delta^{i_{\sigma(n)}}_{j_n} \,/(2^{n-1} (n+2)!),
\label{s5integ} \ee
 where $S_n$ is the permutation group on $n$ objects and we have
normalized $\int_{S^5}1=1$. The index structure is determined by
$SU(3)$ invariance, and the
 normalization factor
 can be determined by multiplying both sides
 by $2^n\, \delta^{i_1}_{j_1} \dots \delta^{i_n}_{j_n}$.\footnote{The
formula, generalized to an arbitrary complex dimension $d, ~1 \le
i_s, j_s \le d, ~1 \le s \le n$ (here $d=3$) is
$\int_{S^{(2d-1)}} z^{i_1} \dots z^{i_n}\, \bz_{j_1} \dots
\bz_{j_n} = \sum_{\sigma \in S_n} \delta^{i_{\sigma(1)}}_{j_1}
\dots \delta^{i_{\sigma(n)}}_{j_n}\, /(2^n c_{n,d})$. The
normalization constant is given by
$c_{n,d}=\sum_{(i_1,\dots,i_n)}^{d^n} \sum_{\sigma \in S_n}^{n!}
\delta^{i_{\sigma(1)}}_{i_1} \cdot \dots \cdot
\delta^{i_{\sigma(n)}}_{i_n}
  =\sum_{i*}^{d^n} n_1 ! \cdot \dots \cdot n_d !
 =\sum_{p*} n!
 = {(n+d-1)! \over n! (d-1)!} n!={ (n+d-1)! \over (d-1)!}$, where
$i*$ is any set of indices $(i_1,\dots,i_n)$ such that there are
$n_1$ 1's and ... $n_d$ d's, and $p*$ is any partition of $n$
objects (the indices) into $d$ bins (the possible values).}

Let us first compute the projection onto the ${\bf 10}$. The
result must be a combination of the ${\bf 10}$s which can be made
out of $h\, h\, \bh$. Actually there are two such combinations
\bea
 {\bf 10}' &=& h_{pq(i} h_{jk)r} \bh^{pqr}, \non
 {\bf 10}''&=& h_{(ijk)} h_{pqr} \bh^{pqr}, \eea
where the brackets () denote symmetrization. We can form a linear
combination of the two from the product of $(h \bh)_{\bf
8}=\bh^{plm}\, h_{klm}-{1 \over 3}\, \delta^p_k \bh^{lmn}\,
h_{lmn}$ with $h$ : \be {\bf 10}''' = [(\bh^{plm}\, h_{klm}-{1
\over 3}\, \delta^p_k\,
  \bh^{lmn}\, h_{lmn})\, h_{pij}+(i j k ~\mbox{perm})]=
  {\bf 10}'-{1 \over 3}{\bf 10}''. \ee
We can express the different terms appearing in the projection
$(ZM{\bf 10},\gammar\, Obst)$ as linear combinations of ${\bf
10}'$ and ${\bf 10}''$ : \bea
 (ZM{\bf 10}, h_j h_k\, \bh^{jk}\, \bz_i\, \gamma^i \gammad_{123}
 \chi_0^*  /r^6) & = & (24 ~{\bf 10}'+0 ~{\bf 10}'')/(8 \cdot 6!), \non
(ZM{\bf 10}, h\, h_{ij}\, \bh^{j}\, \gamma^i \gammad_{123}\,
 \chi_0^*  /r^6) & = & (24 ~{\bf 10}'+0 ~{\bf 10}'')/(8 \cdot 6!), \non
(ZM{\bf 10}, h\, h_{jk}\, \bh^{jk}\, \bz_i\, \gamma^i
\gammad_{123}\,
 \chi_0^*  /r^6) & = & (18 ~{\bf 10}'+6 ~{\bf 10}'')/(8 \cdot 6!), \non
(ZM{\bf 10}, h\, h _{klm}\, \bh^{klm}\, \bz_i\, \gamma^i
\gammad_{123}\, \chi_0^*  /r^4) & = & (0 ~{\bf 10}'+6 ~{\bf
10}'')12/(8 \cdot 6!), \non (ZM{\bf 10}, h_k h_{ilm}\,
\bh^{klm}\, \gamma^i \gammad_{123}\, \chi_0^*  /r^2) & = & (2
~{\bf 10}'+0 ~{\bf 10}'')60/(8 \cdot 6!), \non (ZM{\bf 10}, h_{ik}
h _{lm}\, \bh^{klm}\, \gamma^i \gammad_{123}\, \chi_0^*  /r^2) &
= & (2 ~{\bf 10}'+0 ~{\bf 10}'')60/(8 \cdot 6!).
 \eea
After summing up all the contributions we find
\bea (ZM{\bf 10}, \gammar Obst) =
 {1 \over 8 \cdot 6!}  & \bigg[ & 16(24 ~{\bf
10}' + 0 ~{\bf 10}'') + 8(24 ~{\bf 10}'+0 ~{\bf 10}'') -12(18
~{\bf 10}'+6 ~{\bf 10}'') \non &-& {2 \over 3} 12(0 ~{\bf 10}'+6
~{\bf 10}'') + 60(2 ~{\bf 10}'+0 ~{\bf 10}'') - 60(2 ~{\bf 10}'+0
~{\bf 10}'') \bigg] \non = {12 \over 8 \cdot 6!} & \cdot & (30
~{\bf 10}' -10 ~{\bf 10}'')={30 \cdot 12 \over 8 \cdot 6!} {\bf
10}'''. \eea
 As expected from field theory,
the projection depends solely on ${\bf 10}'''$ ! Thus, it vanishes
if and only if $(h \bh)_{\bf 8}$ vanishes. In the framework of our
computation the fact that the ratio of the two terms is exactly
$(-3)$ is miraculous, though we expect (because of the
correspondence with field theory) that it
``had to happen'' due to a deeper principle which is not manifest
here. In order to write down the projection itself one needs to
multiply by a normalization factor of $4!$, and we find that the
projection is given by
\be\label{obstruction}
 \gammar \, Obst \to
{3 \over 2}\,(\bh^{plm} h_{klm}-{1 \over 3}\, \delta^p_k\,
\bh^{lmn}
 h_{lmn})\, h_{pij}\, z^i z^j\,  \gamma^k  \gammad_{123}\, \chi_0^*/r^3.
\ee

Similarly, we compute the projection on $ZM{\bf 8}$. Here we can
form only one ${\bf 8}$ out of $h\, h\, \bh$ and it has $(h
\bh)_{\bf 8}$ as a factor,
 \be {\bf 8}^i_j=\epsilon^{ikl}\, \bh^{pqr}\, h_{jkp}\,
h_{lqr} = \epsilon^{ikl}\, [(h \bh)_{\bf 8}]^m_k\, h_{jlm}. \ee
There are various terms in $Obst$ that contribute to the
projection,
but we find that the different contributions cancel each other
and we have \be (\gammar ZM{\bf 8},Obst)=0. \ee Thus, as
expected, we find an obstruction to constructing a conformally
invariant solution at third order if and only if $(h \bh)_{\bf
8}$ is non-zero. We expect that when this vanishes we can
construct a conformally invariant solution to all orders, but we
have not been able to prove this explicitly.

\section{Comparison to Field Theory and Non-renormalization}
\label{non-renorm}

In section \ref{perturb} we found that when the combination $(h
\bh)_{\bf 8}$ is non-zero, we have an obstruction to solving the
third order SUSY equations with a conformally invariant solution.
This is exactly what we expect from our field theory analysis in
section 2. In this section we will make the comparison more
precise by comparing the logarithmic running of the coupling
constants in the field theory (at one-loop) and in supergravity.

The obstruction term which we find at third order leads to a
logarithmic term in $\chi^{(3)}$, which using
(\ref{obstruction}),(\ref{obstdef}) is of the form
\be \chi^{(3)}_{log} = -{3\over 2} \log(r/r_0) h_{ijk}^{(3)} z^j
z^k \gamma^i \gamma^{\dagger}_{123} \chi_0^* / r^2, \ee where we
denoted the ``bad" combination ${\bf 10}^{\prime \prime \prime}$
by
\be \label{hthree} 6 h_{ijk}^{(3)} = (\bh^{plm} h_{klm} - {1\over
3} \delta^p_k |h|^2) h_{ijp} + (permutations\ of\ i,j,k). \ee
Using our equation for $\delta \psi$ (\ref{odd2}) we find that
such logarithmic terms in $\chi^{(3)}$ map directly to logarithmic
terms in $G^{(3)}$, of the same form as the original terms we had
in $G^{(1)}$ (except for the logarithmic dependence on $r$). Since
these original terms corresponded to the coupling constants
$h_{ijk}$, it is natural to interpret the logarithmic corrections
as corresponding to a logarithmic running of the coupling
constants $h_{ijk}$. By solving for these terms in $G$, we find
that up to third order the couplings run as
\be\label{running} h_{ijk}(r) = h_{ijk} + 3 h^{(3)}_{ijk}
\log(r/r_0) + \cdots, \ee
for some arbitrary $r_0$. Interpreting $r$ as the scale at which
we measure the running coupling, this is similar to the field
theory expressions which we get from the beta function
(\ref{beta}) and the gamma function (\ref{obstone}). Of course,
the solution we find in which $G$ behaves as in (\ref{running})
cannot be valid for all values of $r$, since eventually the
logarithm becomes large and higher order terms cannot be
neglected; this is similar to what happens in field theory when
the coupling runs. But, we can assume that there is some range of
values of $r$ where the running coupling is small and the
approximation (\ref{running}) is good; this is similar to what we
do in field theory when we have a running coupling, and we impose
UV and IR cutoffs.

To make a precise comparison we need to relate $h$ to the
coupling constant appearing in the field theory superpotential
(namely, to determine its normalization). For this comparison it
will be convenient to return the powers of $\kappa$ and $R_0$
which we have ignored until now; returning these we find that
(\ref{running}) becomes \be\label{nrunning} h_{ijk}(r) = h_{ijk}
+ 3 {\kappa^2 \over R_0^8} h^{(3)}_{ijk} \log(r/r_0) + \cdots. \ee

One possible way to determine the normalization of $h$ is by
carefully normalizing the operators $\tr(\Phi^i \Phi^j \Phi^k)$
on both sides and using the fact that their 2-point function is
not renormalized. We will choose a different approach, which is
to look at a D3-brane probe in the background we computed above,
and to compare the scalar potential on this probe with the scalar
potential we get in the field theory when we turn on a single
non-zero eigenvalue for the matrices $\Phi^i$ (which is
identified with the position of the D3-brane probe). Note that
even though the D3-brane is not static, we can trust the naive
computation of the potential on the brane since we are assuming
that $h$ is very small, and the motion of the D3-brane will be a
higher order effect. Possible corrections to the field theory K\"
ahler potential will also only contribute at higher orders in $h$.

The D3-brane action in the approximation of slowly varying
background fields is of the form \be\label{dthree} S_{D3} =
T_{D3} \int d^4x \sqrt{\det(G_{\mu \nu} + 2\pi \alpha' (B_{\mu
\nu} - F_{\mu \nu}))} - \kappa T_{D3} \int d^4x A_4, \ee where
$T_{D3} = 1 / (g_s (2\pi)^3 (\alpha')^2)$ is the D3-brane
tension, and pull-backs of the bulk fields to the D3-brane
worldvolume are implied everywhere.

First, from the metric term we can read off the kinetic term for
the transverse positions $z^i, \bz_i$ of the D3-brane to zeroth
order in $h$, in the coordinate system we used above. We find
that the kinetic term is precisely given by $T_{D3} \del_{\mu}
z^i \del^{\mu} \bz_i$ (all the powers of $r/R_0$ cancel).

Next, we can compute the potential for the D3-brane, by expanding
the terms in the action with no derivatives to second order in
$h$. By plugging into (\ref{dthree}) the expressions we computed
above, we find that the scalar potential is given by
\be\label{dpot} V_{D3} = T_{D3} {r^4 \over R_0^4} (e^{4\rho} -
f_5) \simeq T_{D3} {r^4\over R_0^4} (4 \rho^{(2)} - \delta
f_5^{(2)}) = T_{D3} {8 \kappa^2 \over R_0^{12}} h_{ijk} z^j z^k
\bh^{ilm} \bz_l \bz_m. \ee

To compare to the field theory we need to set the normalizations
of the field theory Lagrangian. We will take the relevant terms
in the deformed SYM Lagrangian to be of the form
\be\label{kinsym} {1\over g_{YM}^2} \int d^4x \tr(\del_{\mu}
\Phi^i \del^{\mu} \Phi^*_i) + \bigg[ {{N}\over {3 (g_{YM}^2
N)^{3/2}}} \int d^4x d^2\theta H_{ijk} \tr(\Phi^i \Phi^j \Phi^k) +
c.c.\bigg], \ee
where we determined the normalization of the field theory coupling
constant $H$ such that it couples to the chiral primary operator
as normalized in the natural way for the 't Hooft large $N$ limit
\cite{Intriligator:1998ig} (in other words, to have a good large
$N$ limit we need to keep $H_{ijk}$ constant and not $h_{ijk}$;
the relation between the two is determined below). By comparing
the kinetic terms, using $4\pi g_s = g_{YM}^2$, we see that we
should identify $z^i \simeq \sqrt{2} \pi \alpha' \Phi^i$.
Plugging this into (\ref{dpot}) we find that it is given by
\be\label{dpottwo} V_{D3} = {16 \pi^2 \kappa^2 (\alpha')^2 \over
R_0^{12} g_{YM}^2} h_{ijk} \Phi^j \Phi^k \bh^{ilm} \Phi^*_l
\Phi^*_m. \ee
Meanwhile, the field theory potential derived from (\ref{kinsym})
is given by (for large $N$)
\be\label{spottwo} V_{field\ theory} = {1\over {g_{YM}^4 N}}
H_{ijk} \Phi^j \Phi^k {\bar H}^{ilm} \Phi^*_l \Phi^*_m. \ee
By comparing the two expressions we find
\be\label{forh} H_{ijk} = {{4 \pi \kappa \alpha' \sqrt{g_{YM}^2
N}} \over {R_0^6}} h_{ijk}, \ee
so our formula (\ref{nrunning}) above gives
\be\label{nhrunning} H_{ijk}^{SUGRA}(r) = H_{ijk} + {3 R_0^4
\over {16 \pi^2 g_{YM}^2 N (\alpha')^2}} H^{(3)}_{ijk}
\log(r/r_0) = H_{ijk} + {3 \over {16 \pi^2}} H^{(3)}_{ijk}
\log(r/r_0), \ee
where $H^{(3)}$ is related to $H$ in the same way (\ref{hthree})
that $h^{(3)}$ is related to $h$.

We can compare this with the field theory expression, using the
fact that in our normalizations
\be\label{betatwo} \beta_{H_{ijk}} \equiv {\del H_{ijk} \over
{\del \log(\mu)}} = -{1\over 2} (H_{ijl} \gamma^k_l + H_{ilk}
\gamma^l_j + H_{ljk} \gamma^i_l), \ee
and the traceless part of the anomalous dimension matrix is given
by (translating (\ref{obstone}) to the normalizations of
(\ref{kinsym}))
\be\label{gammatwo} \gamma^j_i \equiv {\del \log(Z^j_i) \over
{\del \log(\mu)}} = -{(N^2-4) \over {8 N^2 \pi^2}} H_{ikl} {\bar
H}^{jkl}. \ee
Assuming that the trace of the gamma matrix vanishes this gives
for large $N$
\be\label{frunning} H_{ijk}^{field\ theory}(\mu) = H_{ijk} +
{3\over {16 \pi^2}} H^{(3)}_{ijk} \log(\mu/\mu_0) + \cdots, \ee
where $\mu$ is the renormalization scale.

Identifying the radial direction $r$ with the renormalization
scale $\mu$ via the UV/IR correspondence \cite{Susskind:1998dq},
\footnote{Putting an IR cutoff on the radial direction at some $r_0$, and defining the
couplings by the values of the fields at $r_0$, is believed to be equivalent to
putting a UV cutoff in the field theory at a scale $\mu_0 \propto r_0$ and
defining the couplings at the scale $\mu_0$, though the precise form of this
cutoff in the field theory is not known.}
we find a precise agreement between the weak coupling computation
(in field theory) and the strong coupling computation (in
supergravity) ! This indicates that perhaps the leading terms in
the $h$-perturbation expansion (and in the $1/N$ expansion) in the
beta functions, and therefore also in the anomalous dimensions,
do not depend on $g_{YM}$ (presumably this is not true for higher
order terms in $h$). Apriori the term in $\gamma$ proportional to
$(h \bh)_{\bf 8}$ could have a coefficient with an arbitrary
dependence on $\lambda_{YM} = g_{YM}^2 N$, but it seems that it
is actually a constant (at least in the large $N$ limit).
The fact that the SUGRA result is independent of $g_{YM}$ seems
to follow from the $SL(2,\IR)$ and $U(1)_Y$ symmetries of the
SUGRA action, as discussed in \cite{Intriligator:1998ig}.
However, it is surprising that we find precisely the same
coefficient in the two limits.
This is presumably related to some non-renormalization theorem
for the coefficient of this term in $\gamma$, but we do not know
how to prove this or to relate it to any of the known
non-renormalization theorems in ${\cal N}=4$ SYM.

\acknowledgments

O.A. would like to thank S. Kachru and E. Silverstein for
collaboration on the early stages of this project in 1998
\cite{unpublished}, and to thank the ITP for hospitality and the
participants of the ITP Spring 1998 workshop on dualities in
string theory for useful discussions. We would like to thank J.
Maldacena for collaboration and insight during early stages of
this work, and D. Kutasov, N. Seiberg, and E. Witten for
important discussions. B.K. would also like to thank B. Acharya,
D. Berenstein, E. Diaconescu, M. Gunaydin, A. Hashimoto, Z. Yin
and additional members of the IAS at Princeton for discussions,
and to acknowledge the wonderful hospitality by the Hebrew
University in Jerusalem, the Weizmann Institute, and Stanford
University during stages of this work. O.A. would also like to
thank the Benasque Center for Science, the Aspen Center for
Physics, the Institute for Advanced Studies, and Tel-Aviv
University for hospitality during various stages of this project.
O.A. and B.K. would like to thank the EuroConference in Crete in
September 2000 for enjoyable hospitality.

This work was supported in part by the Einstein Center at the
Weizmann Institute. The work of O.A. and S.Y. is supported in
part by the Israel-U.S. Binational Science Foundation. The work
of O.A. is supported in part by the ISF Centers of Excellence
Program, by the European network HPRN-CT-2000-00122, and by
Minerva. The work of B.K. is supported by DOE under grant no.
DE-FG02-90ER40542, and by a Raymond and Beverly Sackler
Fellowship. The work of S.Y. is supported in part by GIF -- the
German-Israeli Foundation for Scientific Research.

\newpage
\appendix
\section{Finite Groups for $\cM_c$}

We showed in section \ref{FT} that near the $\cn=4$ fixed line,
the space $\cM_c$ \footnote{Throughout this section we will
 use the shorthand notation $\cM_c$ for $\cM_c(\cn=4)$
 and ignore
 the dimension of this space arising from the gauge
coupling constant.} was parameterized by ten couplings $h_{ijk}$
in the $\bf 10$ representation of $SU(3)$, obeying $(h \bh)_{\bf
8}=0$.
 The constraint has just the right form so that the space of
couplings up to global $SU(3)$ transformations is the holomorphic
quotient $\ten/SL(3,\IC)$.
 Here we would like to show that ${\bf
10}/SL(3,\IC)=\IC^2/\cT$, where $\cT = \SLt{\IZ_3}$ is embedded in
$U(2)$ in a way which we describe later, such that it is a double
cover over the tetrahedral group in $SO(3) \simeq U(2)/U(1)$. We
demonstrate this using some discrete subgroups of the $SU(3)$
gauge group, and then we mention how this result would appear
from the usual analysis of gauge invariant coordinates on the
moduli space. We will find that $\IC^2/\cT \simeq \IC^2$ as a
complex variety, namely, $\cM_c$ can be charted by two
unconstrained complex coordinates.

The starting point is to define a discrete subgroup $G_1 \subset
SU(3)$ such that the adjoint (${\bf 8}$) will not have any $G_1$
invariants, which is equivalent to requiring that $G_1$ does not
commute with any generator, or that its normalizer is discrete.
Given such a subgroup and any representation $R$, its $G_1$
invariant subspace $R_{G_1}$ is transverse to the gauge orbits,
and this is helpful in describing $R/SL(3,\IC)$. More precisely,
$R_{G_1}$ after being divided by a residual discrete gauge group
is a subspace of $R/SL(3,\IC)$, and in our case the inclusion
will be shown to be an equality by dimension counting.

We take $G_1$ to be the ``fuzzy torus'' or ``shift and clock''
subgroup defined by two generators $U,\, V$ obeying $U^3=V^3=1, ~
U\, V = \omega\, V\, U\,$ where $\omega=\exp(2 \pi\, i/3)$ is a
third root of unity. It is a group of order 27, with a $Z(G_1)
\simeq \IZ_3$ center generated by $\omega$, and from the defining
relations we see that $G_1/Z(G_1)=\IZ_3 \times \IZ_3$. An explicit
representation of $G_1$ is given by
 \bea
 U &=& \left[
 \begin{array}{ccc}
 0 & 1 & 0 \\
 0 & 0 & 1 \\
 1 & 0 & 0
 \end{array} \right]
\mbox{  -``shift'',} \non
   V &=& \left[
 \begin{array}{ccc}
 1 & 0 & 0 \\
 0 & \omega & 0 \\
 0 & 0 & \omega^2
 \end{array} \right]
 \mbox{  -``clock''.}
 \eea

Actually we can simply ignore the center of $SU(3)$ since both the
$\ten$ and the $\eight$ representations, which we will be
interested in, have zero triality, and thus are representations
of $SU(3)/\IZ_3$. From this point of view ${\tilde G}_1 \equiv
G_1/\IZ_3 \simeq \IZ_3 \times \IZ_3 \subset SU(3)/\IZ_3$.

As anticipated, the ${\bf 8}$ of $SU(3)$ does not have $G_1$
invariants.\footnote{An element $\gamma \in {\bf 8}$, a traceless
3 by 3 matrix, transforms under $G_1$ as $\gamma \to g \gamma
g^{-1}$. Taking $g=V$ limits $\gamma$ to be diagonal, and then
taking $g=U$ leaves $\gamma = a \delta_i^j$ which vanishes due to
the vanishing of the trace.} This is equivalent to saying that
once we impose $G_1$ invariance the $\gamma$ function conditions
simplify to the form $\gamma_i^j = \gamma \delta_i^j$.

Looking for $G_1$ invariants of the ${\bf 10}$ it is convenient
to represent the ${\bf 10}$ by polynomials of degree three. The
only polynomials which are invariant under $G_1$
are\footnote{Invariance under $V$ leaves 4 operators: $\phi_1
\phi_2 \phi_3,~\phi_1^3,~\phi_2^3,~\phi_3^3$. Then, invariance
under $U$ leaves only $O_1, ~O_2$ as stated.}
 \bea
 O_1 &=& \sqrt{6}  \phi_1 \phi_2 \phi_3,
 \non
 O_2 &=& {1 \over \sqrt{3}} (\phi_1^3 + \phi_2^3 + \phi_3^3).
 \eea
We arrive at the same operators as in the field theory analysis,
eq. (\ref{suppot}), while the normalizations here were chosen to
impose unit length in the natural metric of the symmetric tensor
product of $\phi$'s. Moreover, $G_1$ is the maximal subgroup that
keeps $P=\IC^2=\mbox{span}(O_1,O_2)$ fixed.\footnote{For $O_1$ to
be invariant $k$ must be a product of a permutation with a
diagonal matrix, while for $O_2$ to be invariant all three
non-zero entries of $k$ must be cubic roots of unity, which
amounts to the characterization of $G_1$.}
We see that the $\ten$ has $2_\IC$ $G_1$-invariants, which equals
the dimension $10_\IC - 8_\IC$ of the quotient $\ten/SL(3,\IC)$,
and so the invariant space covers all of the quotient, up to
identifications.

In order to find the identifications on the plane $P$, which are
the main claim of this appendix, we note that fixing a
representation of $G_1$ breaks the $SU(3)$ gauge group to the
normalizer $N(G_1) \subset SU(3)$, (namely $N(G_1)$ is the
subgroup that keeps $P$ invariant\footnote{{\it proof:}
$N(G_1)=\{ n \in SU(3) | n ~G_1 ~n^{-1} = G_1 \}$. First suppose
there is a group element $n$ such that $n\, P=P$. Then for all $O
\in P$ $n g n^{-1} O=O$ so $P$ is invariant under $k=n g n^{-1}$,
but $G_1$ is the maximal subgroup that keeps $P$ invariant and so
$k \in G_1$. Conversely, if $n$ is a group element in $N(G_1)$
then for any $g \in G_1$ there is a $g' \in G_1$ such that $g\, n
= n\, g'$. In particular, for all $O \in P$ $g\, n O= n\, g'
O=n\, O$, and so $n O \in P$, namely $n P= P$. QED.} ), and the
identifications are given by the action of $N(G_1)/G_1$.

{\it Claim:} $N(G_1)/G_1=SL(2,\IZ_3)=\cT$.

{\it Proof:} First we will show that $N(G_1)/G_1 \subset
SL(2,\IZ_3)$, and then we will write down its generators and
verify that it actually equals $SL(2,\IZ_3)$.

Any element $n \in N(G_1)/G_1$ is mapped to $SL(2,\IZ_3)$ by the
induced conjugation automorphism of $G_1/\IZ_3=\IZ_3 \times
\IZ_3$, namely an invertible element $M(n)$ of $GL(2,\IZ_3)$.
This automorphism keeps $\omega$ invariant and sends
 \bea U \to U^{M_{11}} ~ V^{M_{12}} ~\omega^{k_1} \\
  V \to U^{M_{21}} ~ V^{M_{22}} ~\omega^{k_2}
 \eea
 where  $M(n)_{ij} \in GL(2,\IZ_3)$ and $k_1, k_2 \in
 \IZ_3$ are some  constants.
The commutator $C=U V U^{-1} V^{-1}=\omega \in G_1$ transforms by
$C \to \omega^{det(M)}$, thus its invariance actually requires $M
\in SL(2,\IZ_3)$. The mapping is actually one to one, namely the
only $n \in N(G_1)$ with $M(n)=1$ are $n \in G_1$. \footnote{Such
an $n$ commutes with $U,V$ and therefore keeps $P$ invariant, and
by the maximal property of $G_1$ we have $n \in G_1$.}

Now let us construct $N(G_1)$ explicitly by adding to $G_1$ the
generators
 \bea
 S={-i \over \sqrt{3}} \left[ \begin{array}{ccc}
 1 & 1 & 1 \\
 1 & \omega & \omega^2 \\
 1 & \omega^2 & \omega
 \end{array} \right],
~~~ T= \left[ \begin{array}{ccc}
 \exp{2\pi i/9} & 0 & 0 \\
 0 & \exp{2\pi i/9} & 0 \\
 0 & 0 & \exp{14\pi i/9}
 \end{array} \right].
 \eea
One can check that $M(S),\,M(T)$ are indeed the usual matrices
$S,T \in \SLt{\IZ_3}$ which are known to generate the whole group,
thereby completing the proof of the claim.

One checks that $P$ is indeed invariant under $S,T$, and that on
$P$ (that is on the column vector $[ O_1 ~O_2 ]$) they are
represented by
 \bea
S_O={i \over \sqrt{3}} \left[ \begin{array}{cc}
 -1 & \sqrt{2} \\
 \sqrt{2} & 1
 \end{array} \right],
~~~ T_O= \left[ \begin{array}{cc}
 1 & 0 \\
 0 & \omega^{-1}
 \end{array} \right].
 \eea

Let us see why $\SLt{\IZ_3}/\IZ_2$ is the tetrahedral group. $\{
1, -1 \}$ is normal in $\SLt{\IZ_3}$, once we divide by it
$\SLt{\IZ_3}/\IZ_2$ can be seen to operate on the projective
space $\IZ_3 {\bf P}^1 \simeq (\IZ_3 - \{0\})/\IZ_2$ which has 4
elements, and therefore $\SLt{\IZ_3}/\IZ_2 \simeq A_4$, the group
of alternating (even) permutations on four elements, which
coincides with the tetrahedral group.

The quotient $\ten/SL(3,\IC)$ can be characterized by invariant
polynomials, namely, polynomials in $h_{ijk} \in \ten$ with
symmetric indices and in the antisymmetric tensor $\eps^{ijk}$.
The lower order invariants can be determined directly -- there
are no invariants of degree 3 or lower, and there is exactly one
of fourth degree,
 \be
h^4= h_{i1,i2,i3}\, h_{j1,j2,j3}\, h_{k1,k2,k3}\, h_{l1,l2,l3}
~\eps^{i1,j1,k1}\, \eps^{i2,j2,l1}\, \eps^{i3,k2,l2}\,
\eps^{j3,k3,l3}. \label{hfour} \ee

In order to carry a full analysis of the invariants and proceed
to higher invariants, we use the relation $\ten/SL(3,\IC)=\IC^2/
\cT$ to reduce the problem to that of finding $\cT$ invariants on
symmetric products of $P$, namely $\mbox{Sym}^k (P)$. The number
of invariants in a representation is given by the character
formula \be \mbox{Inv}={1 \over |\cT|} \sum_{g \in \cT} \chi(g),
\ee where $\chi(g)=\mbox{tr}(g)$, and the order of the group is
$|\cT|=24$. Since the character is constant over conjugacy
classes it is useful to determine those to be \be
\begin{array}{|c|ccccccc|}
\hline
\mbox{element}      & {\bf 1} & {\bf -1} & S,\,S^{-1} & T & T^2 & -T & -T^2 \\
\hline
\mbox{multiplicity} & 1       & 1        & 6          & 4 & 4   & 4  & 4 \\
\mbox{ diagonal form }& (1,1)   & (-1,-1)  &(i,-i)      &
(1,\omega)& (1,\omega^2)&
(-1,-\omega)& (-1,-\omega^2) \\
\hline
\end{array} \ee
where the diagonal form refers to the diagonalized form of our 2d
representation. One finds that the number of invariants of
$\mbox{Sym}^k (P)$ vanishes for odd $k$, while for even $k=2\, j$
it is $\mbox{Inv}(j)=j/6+\left[ 2+(-)^j 6+ (-)^j 16\, \cos(2 \pi
j/6) [\cos (2 \pi j/3) + \sin(2\pi j/3)/\sqrt{3} ] \right]/24$.
Once the number of invariants is tabulated one finds that the
generating function is given by $\sum \mbox{Inv}(j)\, q^j=
1/[(1-q^2)(1-q^3)]$, which means that there are two and only two
primitive invariants -- one at order $k=2j=4$ (which we described
directly in (\ref{hfour})) and one at order $k=2j=6$.

Going back to the $SL(3,\IC)$
invariants this implies the existence of a second invariant
$h^6$. So there are exactly 2 invariants (coordinates) on this
space
 \be h^4,\, h^6, \ee
 with no relations among them. Therefore
the space is not singular as a complex variety, $\IC^2/\cT \simeq
\IC^2$, though it may have a metric singularity. It would be
interesting to derive this directly in terms of
$SL(3,\IC)$
 invariants rather than using $\cT$.

\section{Some $SU(3)$ and $SU(4)$ Representations}

In this paper we denote the irreducible representation by their
dimension, and when ambiguous we use the conventions of
\cite{slansky}. We list here a dictionary of the Dynkin indices
(which are directly related to the Young tableaux) for these
representations, together with some decompositions. The Dynkin
indices are given in parentheses and the commas between them are
omitted, as is customary.

\vspace{.5cm} \underline{$SU(3)$ representations :} \be
\begin{array}{ccc p{2cm} ccc}
{\bf 3} &=& (10) & &{\bf 15'}&=& (40) \\
{\bf 6} &=& (20) & &{\bf 21} &=& (05) \\
{\bf 8} &=& (11) & &{\bf 24} &=& (13) \\
{\bf 10} &=& (30) & &{\bf 27} &=& (22) \\
{\bf 15} &=& (21) & &{\bf 28} &=& (60)
\end{array}
\ee

\vspace{.5 cm} \underline{$SU(4)$ representations and useful
decompositions under $SO(6) \to SU(3) \times U(1)$} :
 \bea
{\bf  4}=(100) & \to & {\bf 3}_1 + {\bf 1}_{-3} \non {\bf
\bar{4}}=(001) & \to & {\bf \bar{3}}_{-1} + {\bf 1}_3 \mbox{ (by
complex conjugation)} \non {\bf  6}=(010) & \to & {\bf
3}_{-2}+{\bf \bar{3}}_2 \non {\bf  10}=(200) & \to & {\bf 6}_2 +
{\bf 3}_{-2} + {\bf 1}_{-6} \non {\bf  15}=(101) & \to & {\bf
8}_0 + {\bf 3}_4 + {\bf \bar{3}}_{-4} + {\bf 1}_0 \non {\bf
45}=(210) & \to & {\bf 15}_4 + {\bf 10}_0 + {\bf 8}_0 + {\bf
6}_{-4} + {\bf 3}_{-8} + {\bf \bar{3}}_{-4} \non {\bf
\overline{20}}=(110) & \to & {\bf 8}_3 + {\bf 6}_{-1} +{\bf
3}_{-5} + {\bf \bar{3}}_{-1} \non {\bf  20'}=(020) & \to & {\bf
6}_{-4} + {\bf \bar{6}}_4 +{\bf  8}_0 \non {\bf
\overline{20''}}=(300) & \to & {\bf 10}_3+{\bf 6}_{-1} + {\bf
3}_{-5} + {\bf 1}_{-9} \non {\bf  36}=(201) & \to & {\bf 15}_1 +
{\bf 8}_{-3} + {\bf 6}_5 + {\bf 3}_1 + {\bf \bar{3}}_{-7} + {\bf
1}_{-3}
 \non
{\bf  60}=(120) & \to & {\bf \overline{15}}_{5}+{\bf 15}_1+ {\bf
10}_{-3}+{\bf 8}_{-3}+{\bf 6}_{-7}+{\bf \bar{6}}_1
 \non
{\bf  64}=(111) & \to & {\bf 15}_{-2} + {\bf \overline{15}}_2 +
{\bf 8}_6 + {\bf 8}_{-6} + {\bf 6}_2 +
 {\bf \bar{6}}_{-2} + {\bf 3}_{-2} + {\bf \bar{3}}_2 \non
{\bf   84}=(202) & \to & {\bf (27+8+1)}_0 + {\bf 15}_4 + {\bf
\overline{15}}_{-4} + {\bf 6}_8 +
 {\bf \bar{6}}_{-8}+ {\bf 3}_4 + {\bf \bar{3}}_{-4} \non
{\bf  84'}=(310) & \to & {\bf 15'}_1 + {\bf 24}_5 + {\bf 15}_1+
{\bf 10}_{-3}+ {\bf 8}_{-3}+{\bf 6}_{-7} +{\bf 3}_1 + {\bf
\bar{3}}_{-7} \non {\bf 105}=(040) & \to & {\bf 15'}_{-8} + {\bf
\overline{24}}_{-4} + {\bf 27}_{0} + {\bf 24}_{4} + {\bf
\overline{15'}}_{8} \non {\bf  126}=(220) & \to & {\bf 27}_6 +
{\bf 24}_2 + {\bf 15'}_{-2} + {\bf \overline{15}}_2 + {\bf
15}_{-2} +
 {\bf 10}_{-6} + {\bf 8}_{-6} + {\bf 6}_{-10} + {\bf \bar{6}}_{-2}
\eea To get the $U(1)$ charge normalization of \cite{slansky},
charges must be multiplied by $(-1/3)$.

\bibliographystyle{JHEP}
\bibliography{conformal}

\end{document}